\documentclass[aps, preprint, superscriptaddress,nobibnotes]{revtex4-2}
\usepackage{color}
\usepackage{graphicx}
\usepackage{epstopdf}
\usepackage{caption}
\usepackage{subcaption}
\usepackage{physics}
\usepackage{xcolor}
\usepackage{bm}
\usepackage{amsmath, amsthm, amssymb, amsfonts}
\usepackage{mathtools}
\usepackage{array}

\usepackage{changes}

\usepackage{xr}
\newcommand{\be}{\begin{equation}}
\newcommand{\ee}{\end{equation}}

\begin{document}
\title{ Multi-orbital Kondo screening in strongly correlated polyradical nanographenes. 
}

\newcommand{\FISI}[0]{{Departamento de F\'{\i}sica, Universidad del Pa\'{\i}s Vasco UPV-EHU, 48080 Leioa, Spain}}
\newcommand{\QUIMI}[0]{{Departamento de Polímeros y Materiales Avanzados: F\'{\i}sica, Qu\'{\i}mica y Tecnolog\'{\i}a, Universidad del Pa\'{\i}s Vasco UPV-EHU, 20018 Donostia-San Sebasti\'an, Spain}}
\newcommand{\DIPC}[0]{{Donostia International Physics Center (DIPC),
20018 Donostia-San Sebasti\'an, Spain}}
\newcommand{\EHUQC}[0]{{EHU Quantum Center, Universidad del Pa\'{\i}s Vasco UPV-EHU, 48080 Leioa, Spain}}
\newcommand{\CFM}[0]{{Centro de F\'{\i}sica de Materiales CFM/MPC (CSIC-UPV/EHU), 20018 Donostia-San Sebasti\'an, Spain}}

\author{Aitor Calvo-Fern{\'a}ndez}
\thanks{These authors contributed equally to this work.}
\affiliation{\FISI}
\affiliation{\DIPC} 

\author{Diego Soler-Polo}
\thanks{These authors contributed equally to this work.}
\affiliation{Institute of Physics, Academy of Sciences of the Czech Republic, Cukrovarnicka 10, Prague 6, CZ 16200, Czech Republic}

\author{Andr{\'e}s Pinar Sol{\'e}}
\thanks{These authors contributed equally to this work.}
\affiliation{Institute of Physics, Academy of Sciences of the Czech Republic, Cukrovarnicka 10, Prague 6, CZ 16200, Czech Republic}

\author{Shaotang Song}
\thanks{These authors contributed equally to this work.}
\affiliation{Department of Chemistry, National University of Singapore, Singapore 117543, Singapore}

\author{Oleksander Stetsovych}
\affiliation{Institute of Physics, Academy of Sciences of the Czech Republic, Cukrovarnicka 10, Prague 6, CZ 16200, Czech Republic}

\author{Manish Kumar}

\affiliation{Institute of Physics, Academy of Sciences of the Czech Republic, Cukrovarnicka 10, Prague 6, CZ 16200, Czech Republic}

\author{Guangwu Li}
\affiliation{Department of Chemistry, National University of Singapore, Singapore 117543, Singapore}

\author{Jishan Wu}
\affiliation{Department of Chemistry, National University of Singapore, Singapore 117543, Singapore}

\author{Jiong Lu}
\affiliation{Department of Chemistry, National University of Singapore, Singapore 117543, Singapore}

\affiliation{Institute for Functional Intelligent Materials, National University of Singapore, 117544,
Singapore}

\author{Asier Eiguren}
\affiliation{\FISI}
\affiliation{\DIPC}
\affiliation{\EHUQC}

\author{Mar{\'i}a Blanco-Rey}
\thanks{Corresponding author}
\affiliation{\QUIMI}
\affiliation{\DIPC}
\affiliation{\CFM}

\author{Pavel Jel\'{\i}nek}
\thanks{Corresponding author}
\affiliation{Institute of Physics, Academy of Sciences of the Czech Republic, Prague, Czech Republic}

\affiliation{Regional Centre of Advanced Technologies and Materials, Czech Advanced Technology and Research Institute (CATRIN), Palacký University Olomouc,Olomouc 78371, Czech Republic}

\date{\today}

\begin{abstract}	
We discuss coexistence of Kondo and spin excitation signals in tunneling spectroscopy in
strongly correlated polyradical $\pi$-magnetic nanographenes on a metal surface. The Kondo signal is rationalized 
by a multi-orbital Kondo screening of the unpaired electrons.
The fundamental processes are spin-flips of antiferromagnetic (AFM) order
involving charged molecular multiplets.
We introduce a~perturbative model, which provides simple rules to identify the presence of AFM channels responsible for Kondo screening. The Kondo regime is confirmed by numerical renormalization group calculations.
This framework can be applied to similar strongly correlated open-shell systems.
\end{abstract}

\maketitle


 
%
   

The molecular $\pi$-magnetism\cite{bib:ovchinnikov78,bib:lieb89,rossier_mirror}  has attracted the attention of researchers in various fields, including chemistry, physics, and materials science \cite{bib:oteyza22,Yazyev2010}. Understanding the nature of $\pi$-magnetism and its underlying principles is crucial for the development of new materials and technologies, such as organic spintronics and molecular electronics \cite{Pi-Magnet2001}. Recent progress in on-surface synthesis \cite{Grill2007,Clair2019} combined with the unprecedented spatial resolution of scanning probe microscopy \cite{Gross2009,Jelnek2017,Ormaza2017} has prompted new studies of $\pi$-magnetic systems \cite{bib:pavlicek17,Mishra2019,Ruffieux2016,bib:song21}. 

From this point of view, one of the interesting magnetic phenomena is the observation of the Kondo effect \cite{bib:hewson,kondo1964resistance}  in open-shell hydrocarbon molecules \cite{Li2019,bib:li20,bib:mishra20,bib:jacob21,bib:turco23}. Originally, the Kondo effect was observed in adatoms with partially occupied $d$-electrons  on metallic substrates \cite{bib:madhavan98,bib:li98,bib:knorr02,bib:otte08}. In principle, the unpaired electrons in extended molecular orbitals can also act as a magnetic impurity and scatter the conduction electrons of the underlying metal surface, leading to the molecular Kondo effect \cite{Esat2015,bib:scott10,bib:komeda11,bib:minamitani12,bib:hiraoka17,bib:zitko21,bib:fernandez08,bib:choi10,bib:perera10,bib:requist14,bib:minamitani15,bib:patera19,bib:koshida21,bib:lu21}. 

While the conventional $S=1/2$ Kondo effect for  magnetic impurities has been extensively studied and is very well understood, the mechanism of a multi-orbital Kondo effect of magnetic impurities with various strongly correlated unpaired electrons is barely explored. In particular, here the so-called underscreened Kondo effect can take place if there are less than $2S$ screening channels \cite{bib:nozieres80}, whereby the ground state shows a residual magnetic moment. This effect was extensively studied for quantum dot systems both experimentally \cite{bib:roch09,bib:parks10} and theoretically \cite{bib:pustilnik01}. Recently, the Kondo effect was also reported for open-shell triplet ($S=1$) nanographenes \cite{bib:li20,bib:turco23} and porphyrins \cite{bib:girovsky2017,yo8,Sun2020} on metallic surfaces. The former case  was theoretically rationalized within the one-crossing approximation \cite{bib:jacob21}. Nevertheless, a broader picture of the multi-orbital Kondo effect in open-shell molecular systems is still missing. Models need to account, on the same footing, for aspects such as the number and symmetry of screening channels or the multiplet nature (i.e. many-electron) of the impurity electronic configuration. Therefore, the emergence of the Kondo resonances in open-shell nanographenes with several strongly correlated unpaired electrons on metallic surfaces provides a~convenient platform for testing such general model.

\begin{figure*}[!htbp]
{\includegraphics[width = 7in]{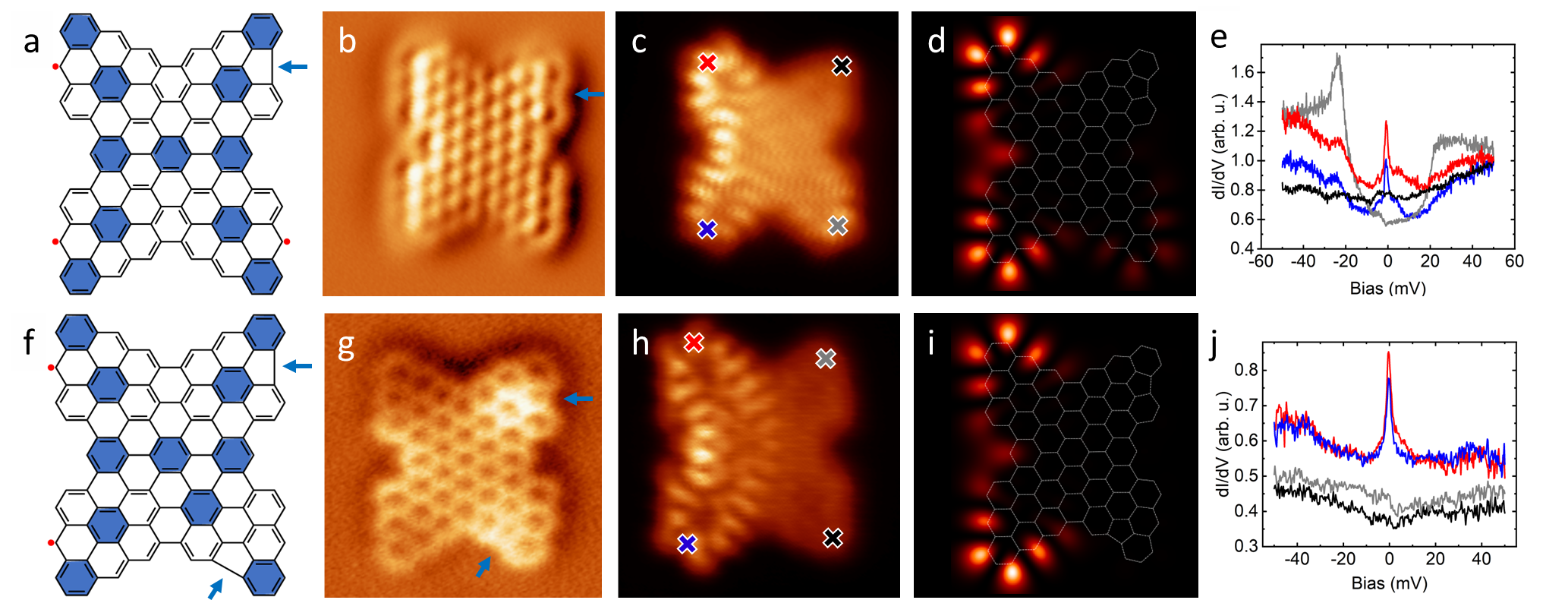}} \qquad
\caption{Panels (a,f) display chemical structure of $\bf 1$ and $\bf 2$, blue arrows highlight the location of the defects. Simultaneously recorded (b,g) high resolution CO tip atomic force microscopy and (c,h) corresponding STM images of the molecules acquired at 1 meV. (d,i) simulated dI/dV maps from the NTO for the spin-flip process between two degenerate ground states, (e,j) dI/dV spectra measured at the positions indicated by crosses on panels (c,h). The Kondo temperature $T_{K}$ is estimated to be approx. 9.5K and 7.5K, respectively, using the Frota function \cite{Prser2012}.  Scanning range of the images: (b,c) 3 $\times$ 3 nm$^2$, (g,h) 2.5 $\times$ 2.5 nm$^2$.}
\label{fig:exp}
\end{figure*}

In this work, we present a theoretical and experimental study of the multi-orbital Kondo effect in strongly correlated polyradical butterfly-shaped nanographenes deposited on a Au(111) surface.  Interestingly, tunneling spectroscopy shows that the Kondo signal coexists with spin-excitation signals within the same nanographene (see Fig.~\ref{fig:exp}).  We build a three-step theoretical framework to explain this behaviour: (i) complete active space configuration interaction (CASCI) calculations \cite{CAS} accurately provide the nanographene electronic states; (ii) a generalized perturbative analysis of the Anderson Hamiltonian, which accounts for the multiplet structure, reveals the antiferromagnetically coupled channels responsible for the Kondo screening by the metal substrate; and finally (iii) numerical renormalization group (NRG) calculations \cite{krishna-murthy1980} provide observable quantities. We propose that the spatially-delocalized Kondo physics is ubiquitous in multireference systems (i.e. those that cannot be entirely described by a single Slater determinant) showing $\pi$-magnetism and can be tackled by our approach.


Recently, we reported on-surface synthesis of a~tetraradical nanographene with a strongly correlated singlet ground state on Au(111) surface (for details see Ref.~\cite{bib:song23}). As side products of the reaction, we frequently found defective nanographenes, which arises from the detachment of  methyl groups during cyclodehydrogenation reaction, leading to 1 or 2 pentagon rings on zig-zag edges (see Fig.~\ref{fig:exp}). In the rest of the manuscript, we will refer as {\bf 1} (Fig.~\ref{fig:exp}a) and {\bf 2} (Fig.~\ref{fig:exp}b), respectively, to the nanographenes with one and two pentagon defects. Importantly, the presence of these defects significantly affects their electronic structure. Namely, we observe zero-energy resonances (see Fig.~\ref{fig:exp}e,j), which are localized on the defect-free zig-zag edge (see Fig.~\ref{fig:exp}c,h). To prove the magnetic origin of the zero-energy resonance we employed a nickelocene functionalized probe ($S=1$) \cite{Ormaza2017}.  Fig.~\ref{SM-fig:IETS} of the Supplemental Material (SM) \cite{bib:SM} shows the strong renormalization of the spin-excitation signal in scanning tunneling spectroscopy (STS) spectra acquired along the tip approach, revealing the magnetic character of {\bf 1} and {\bf 2} \cite{bib:song23}. Thus, we assign the zero-energy resonances to the Kondo effect.
 
Interestingly, in the case of $\bf 1$, the STS spectra acquired at different positions of the molecule reveal a coexistence of an inelastic signal at 25 meV and a zero-bias Kondo resonance within the same molecule (see Fig.~\ref{fig:exp}e). The inelastic signal, tentatively attributed to a spin excitation from the ground state to the first excited state, is located on a defect-free triangulene corner opposite to the zig-zag edge hosting the Kondo resonance (see Figs.~\ref{SM-fig:IETS} and \ref{SM-fig:exp-SM-excDQ}). 

To understand the electronic structure of $\bf 1$ and $\bf 2$ we employ the Hubbard model:
 \begin{equation}
 \label{Hubbard}
      \hat{H} = t \sum_{\langle \mu,\nu  \rangle, \sigma } \hat{c}^\dagger_{\mu \sigma}\hat{c}_{\nu \sigma} + \text{h.c.} + U \sum_{\mu} \hat{n}_{\mu \uparrow} \hat{n}_{\mu \downarrow}.
 \end{equation}
 where $\hat{c}^\dagger_{\mu \sigma}$ ($\hat{c}_{\mu \sigma}$) are creation (annihilation) operators associated to the $\mu$-labelled L\"owdin $\pi$-orbitals on a carbon atomic site with $\sigma$ spin index, $t$ represents hopping between the nearest atomic sites, and $U$ is the repulsive on-site Coulomb interaction introducing many-body electron-electron interaction in the system. We used values $t$ = -2.8 eV and $U$ = 4.3 eV,  as parameterized for an effective Hubbard model in $\pi$-conjugated hydrocarbons \cite{hubbard_optimal_graphene}.

\begin{figure}[!htbp]
{\includegraphics[width=1.0\linewidth]{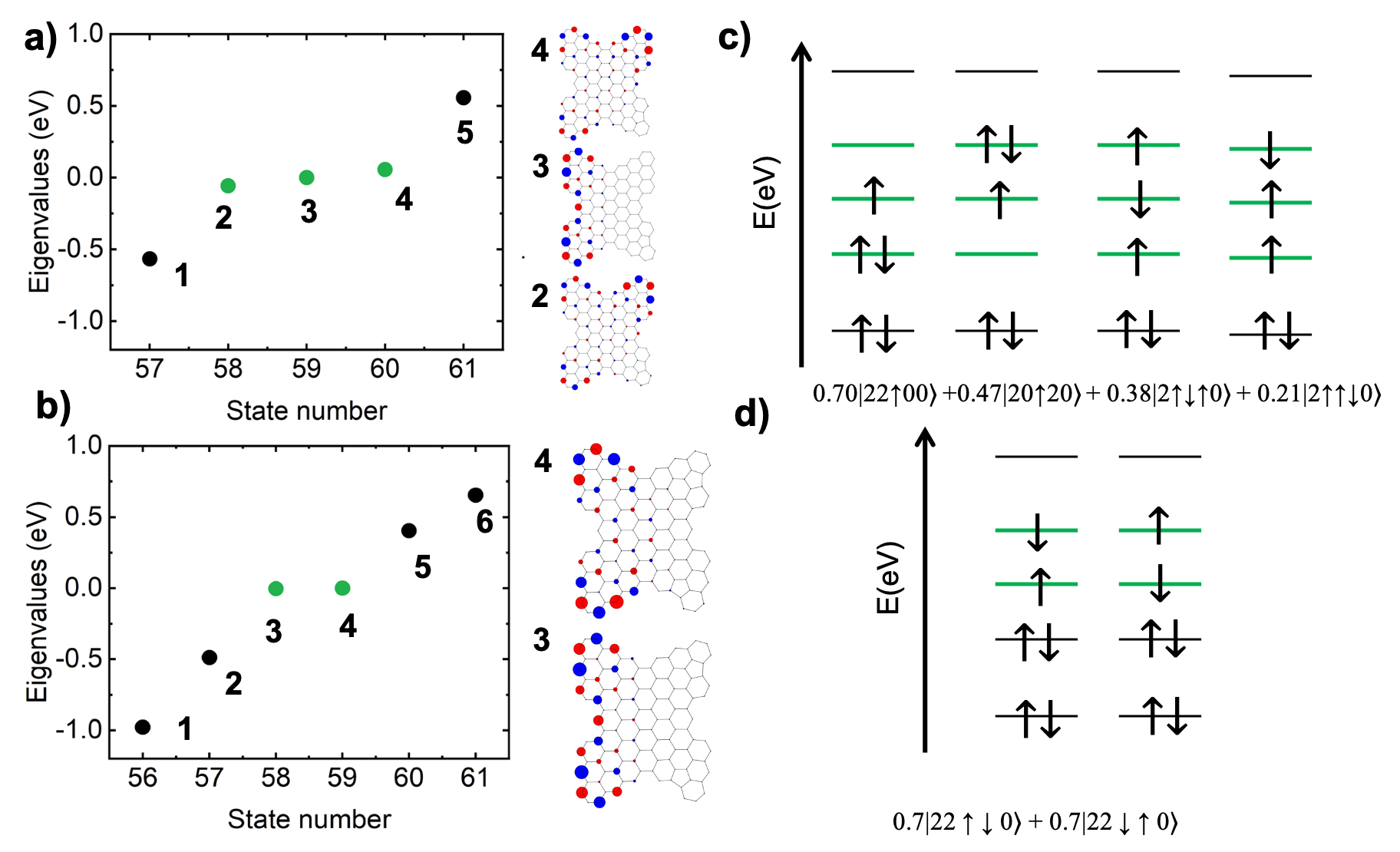}}
\caption{Panels (a,b) show the single electron H{\"u}ckel energy spectrum and the corresponding frontiers orbitals of \textbf{1} and \textbf{2}, respectively. In green we highlight the orbitals that are considered later in the NRG calculations. 
Panels (c,d) display a scheme of the multi-referential ground state obtained from Hubbard-CASCI(5,5) (resp. CASCI(6,6)) for \textbf{1} and \textbf{2}. For more details about the electronic structure, see discussion in SM \cite{bib:SM}.} 
\label{fig:CAS}
\end{figure}

Given the multireferential character of these systems, we solve the Hamiltonian eq.~\ref{Hubbard} by CASCI calculations in active spaces of up to six orbitals \cite{CAS}.  Fig.~\ref{fig:CAS} displays the one-electron energy spectrum and corresponding Huckel orbitals selected for the spaces.

According to Hubbard CASCI(5,5) calculations $\bf 1$ has a~triradical duplet ($S=1/2$) ground state. Importantly, we observe that both the ground and excited states have strong multireferential character with contribution of several non-negligible Slater determinants (see Fig.~\ref{fig:CAS}c). We find that the first excited quartet state is 28 meV \cite{bib:SM} higher than the ground state, which matches very well the experimentally observed spin excitation at 25 meV.  

To rationalize the spatial distribution of the differential
low-bias STM images shown in Fig.~\ref{fig:exp}c,h, we
calculate the natural transition orbitals \cite{NTOs} (NTO)
corresponding to distinct spin-flip processes. In
particular, we calculated the NTOs corresponding to the
spin-flip process between two degenerate ground doublet
states ($\vert M_{0},\pm1/2 \rangle$), see
Fig.~\ref{SM-fig:NTO_B3_doublet2doublet}.  The simulated STM
image \cite{Krej2017} from NTO shown in Fig.~\ref{fig:exp}d
fits nicely to the experiment (Fig.~\ref{fig:exp}c).
Similarly, we also calculated the NTO corresponding to the
spin excitation from the  doublet ground state $\vert
M_{0},\pm 1/2 \rangle$ and the first excited quartet state
$\vert M_{1} \rangle$.
Fig.~\ref{SM-fig:NTO_B3_doublet2quartet} displays the
calculated dI/dV maps of the corresponding NTO orbitals
which match well to the experimental dI/dV map acquired at
24 meV (Fig. ~\ref{SM-fig:exp-SM-excDQ}). 

In the case of $\bf 2$, the ground state is a multi-reference diradical triplet ($S=1$) according to the Hubbard model solved using CASCI(6,6). It is almost exclusively formed by two determinants, with different occupation of singly-unoccupied/unoccupied molecular orbitals (see Fig.~\ref{fig:CAS}d)). Fig.~\ref{fig:exp}i displays calculated dI/dV map of NTOs corresponding to spin flip between two degenerated triplet states ($\vert M_{0}, \pm 1 \rangle $ and $\vert M_{0}, 0 \rangle$), which matches very well to the experimental dI/dV maps at zero bias, as shown in Fig.~\ref{fig:exp}h. 

One way to rationalize the presence of the Kondo effect could be the presence of an uncorrelated unpaired electron. In principle, this uncorrelated single electron could act as a $S=1/2$ magnetic impurity with Kondo screening mechanism by the bath of free electrons in the metallic surface.  Thus, we evaluated the spin correlations, $\langle \hat{\Vec{S}}_j \cdot \hat{\Vec{S}}_k  \rangle - \langle \hat{\Vec{S}}_j \rangle  \cdot \langle \hat{\Vec{S}}_k  \rangle $, between the three unpaired electrons in the doublet ground state (see SM \cite{bib:SM}).  The values in the Tables~\ref{SM-spincorr1} and \ref{SM-spincorr2} reveal strong entanglement between all unpaired electrons of {\bf 1} and {\bf 2}. Thus the presence of the Kondo effect cannot be explained within the traditional framework of a $S=1/2$ magnetic impurity screened by a bath of free electrons of the metal.


The emergence of Kondo peaks is explained here with the ionic multichannel Anderson Hamiltonian \cite{anderson1970}, which is converted into a Kondo Hamiltonian by application of a Schrieffer-Wolf canonical transformation \cite{schrieffer1966,flores2017}. This filters out high-energy fluctuations, keeping second order scattering processes. The starting molecule-substrate hybridization contribution has the form
\begin{align}
\label{AndersonHamiltonian}
\hat{H}_\text{hyb}
    =
    \int_{-D}^{D}
    d\epsilon 
   \rho^\frac{1}{2}
    V
    \bra{M_f m_f}
    \hat f^\dagger_{\alpha \sigma} \ket{M_i m_i}
   \ket{M_f m_f}\bra{M_i m_i}
    \hat{c}_{\epsilon \alpha \sigma}
    + \text{h.c.},
\end{align}
where $V$ is the hybridization amplitude, repeated indices are summed over and the integral is made over a conduction band of width $2D$ and constant density of states $\rho$. 
Since the molecular orbitals (MO) are orthogonal and their substrate-mediated hybridization is small, we assume that each channel couples to one specific CASCI active space MO, both labeled $\alpha$. 
\footnote{Since the MOs are orthogonal and their substrate-mediated hybridization is small, we consider the approximation that each channel couples to one specific MO only.}. 
Each term introduces electron-hole pair excitations amounting to creation (annihilation) of an electron in the MO $\alpha$ by the operator $\hat f^\dagger_{\alpha \sigma}$ ($\hat f_{\alpha \sigma}$) and annihilation (creation) of a conduction electron by the operator $\hat c_{\epsilon\alpha\sigma}$ ($\hat c^\dagger_{\epsilon\alpha\sigma}$).
The multiplet states are labeled $\ket{Mm}$. Aside from the ground multiplet $M_{0}$, we keep only the charged multiplets $M_c$ with the lowest energy, which are obtained also from CASCI calculations by removing or adding one electron within the same active space. These are the main contributions to the molecule Kondo physics (explicit expressions are given in SM section~\ref{SM-sec:IAH}\cite{bib:SM}).


The Kondo Hamiltonian resulting from the canonical transformation is
\begin{equation}
    \hat H_\text{K}
    =
    \frac{J_{\alpha\alpha'}}{2}
    \hat{\mathbf{S}} 
    \cdot
    \hat{\mathbf{s}}_{\alpha\alpha'}
\end{equation}
where $J_{\alpha\alpha'}$ are the effective coupling constants, $\hat{\mathbf S}$ is the molecular spin operator, and $\hat{\mathbf{s}}_{\alpha\alpha'}$ are the conduction spin density matrix elements at the locations of the MOs $\alpha$ and $\alpha'$
\footnote{We have neglected the potential scattering term,  as it does not affect the spin of the molecule. }.
The obtained expressions for $J_{\alpha\alpha'}$ are (see SM section~\ref{SM-sec:Kondo}\cite{bib:SM})
\begin{equation}
    J_{\alpha\alpha'}
    = 
    \sum_{M_c}
    \frac{V^2}{\epsilon_{M_c}-\epsilon_{M_0}}
    C^{M_c}_{\alpha\alpha'},
\label{eq:J}
\end{equation}
\begin{equation}
    C^{M_c}_{\alpha\alpha'}
    =
    \begin{cases}
    \frac{
        \bra{M_0}|\hat{f}_{\alpha}|\ket{M_c}
        \bra{M_c}|\hat{f}^\dagger_{\alpha^\prime}|\ket{M_0}
    }{
        (S_0+\frac{1}{2})
    }, & N_0+1,S_0-\frac{1}{2},
    \\
    \frac{
        \bra{M_0}|\hat{f}^{\dagger}_{\alpha^\prime}|\ket{M_c}^*
        \bra{M_c}|\hat{f}_{\alpha}|\ket{M_0}^*
    }{
        S_0
    }, & N_0-1,S_0-\frac{1}{2},
    \\
    - \frac{
        \bra{M_0}|\hat{f}_{\alpha}|\ket{M_c}
        \bra{M_c}|\hat{f}^\dagger_{\alpha^\prime}|\ket{M_0}
    }{
        (S_0+\frac{1}{2})
    }, & N_0+1,S_0+\frac{1}{2},
    \\
    - \frac{
        \bra{M_0}|\hat{f}^{\dagger}_{\alpha^\prime}|\ket{M_c}^*
        \bra{M_c}|\hat{f}_{\alpha}|\ket{M_0}^*
    }{
        (S_0+1)
    }, & N_0-1,S_0+\frac{1}{2}.
    \end{cases}
\label{eq:C}
\end{equation}
where each coefficient $C_{\alpha\alpha'}^{M_c}$ is determined by the number of particles ($N_0\pm1$) and the spin ($S_0\pm\frac{1}{2}$) of the excited multiplet $M_c$.

The coupling constants $J_{\alpha\alpha'}$ form a hermitian matrix $\mathbf J$ that can be diagonalized to rewrite the effective interaction as $H_K = j_a \hat{\mathbf{S}} \cdot \hat{\mathbf{s}}_{aa}$, where $j_a$ are the eigenvalues and $\ket a = U_{\alpha a}\ket{\alpha}$ the corresponding rotated MOs, which we denote {\it Kondo orbitals}.
$j_a>0$ means that the coupling is antiferromagnetic (AFM) and $j_a<0$ ferromagnetic (FM). The latter does not qualitatively alter the low-temperature physics of the system. Therefore, the regions where Kondo effects are observed are those where Kondo orbitals $\ket{a}$ such that $j_a>0$ are localized, offering a rigorous alternative to NTOs in the interpretation of dI/dV maps (see SM section~\ref{SM-subsec:NTO-KO} and Figs.~\ref{SM-fig:compare_NTO_KO_B3} and \ref{SM-fig:compare_NTO_KO_B2}). Thus expressions in eq.\ref{eq:C} represent a simple rule to identify AFM channels responsible for Kondo screening.

\begin{table}
\begin{tabular}{ r | m{1.5cm} m{1.5cm} m{1.5cm} m{1.5cm} }
    \hline
    $M_c$
    &$M_{0-}$
    &$M_{1-}$
    &$M_{0+}$
    &$M_{1+}$
    \\
    \hline
    2 & 0.86 & 0 & -0.53 & 0 \\
    $\alpha\qquad$ 3 & 0 & 0.75 & -0.07 & -0.71 \\
    4 & -0.26 & 0 & -0.63 & 0 \\ 
    \hline
\end{tabular}
\\
\begin{tabular}{ r | m{1.5cm} m{1.5cm} m{1.5cm} m{1.5cm} }
    \hline
    $M_c$
    &$M_{0-}$
    &$M_{1-}$
    &$M_{0+}$
    &$M_{1+}$
    \\
    \hline
    $\alpha\qquad$ 3 & 0.77 & 0.80 & 0.95 & 0.98\\
    4 & -0.52 & 0.53 & 0.63 & 0.65 \\
    \hline
\end{tabular}
\caption{\label{tab:contributions}Reduced matrix elements $\bra{M_c}|\hat{f}^{(\dagger)}_\alpha|\ket{M_0}$
    for positively (+) and negatively (-) charged excited multiplets $M_c$ for $\bf 1$ (top table) and $\bf 2$ (bottom table).} 
\end{table}


For $\bf 1$ we use a 3-channel Anderson model, with each channel coupled to one of the 3 MOs highlighted in green in Fig. \ref{fig:CAS}, labeled $\alpha=2,3,4$, since these are the only ones with a sizeable magnetic coupling to the molecule. The necessary data for computing the coupling constants for these channels are given in Table~\ref{tab:contributions}. The energies $\epsilon_M$ have been approximated as $\epsilon_-=\epsilon_{M_{0-}}=\epsilon_{M_{1-}}$ and $\epsilon_+=\epsilon_{M_{0+}}=\epsilon_{M_{1+}}$. Using eq. \ref{eq:J}, the matrix $\bf J$ is, up to two digits of precision in the coefficients,
\begin{equation}
    \mathbf{J}
    =
    V^2
    \begin{pmatrix}
        -\frac{0.40}{\epsilon_-} - \frac{0.29}{\epsilon_+} & -\frac{0.06}{\epsilon_+} & 0 \\
        -\frac{0.06}{\epsilon_-} & \frac{1.13}{\epsilon_-}+\frac{0.51}{\epsilon_+} & 0 \\
        0 & 0 & -\frac{0.04}{\epsilon_{-}}-\frac{0.4}{\epsilon_+} 
    \end{pmatrix}
\end{equation}
The only positive eigenvalue $j_a$ couples the molecular spin primarily to channel $\alpha=3$, with a small mixing $|U_{23}|^2\sim 0.001$ with $\alpha=2$. Therefore, we predict a Kondo resonance to be located primarily in the orbital 3 region. This is in agreement with the experimental results, which show a strong Kondo signal in the left wing of the molecule, where the Hückel orbital 3 is located (see Figs. \ref{fig:exp} and \ref{fig:CAS}).

For $\bf 2$, the channels contributing to the magnetic coupling are those highlighted in green in Fig. \ref{fig:CAS}, labeled $\alpha=3,4$. Following the same procedure as for $\bf 1$, we use the data in  Table~\ref{tab:contributions} to obtain
\begin{equation}
    \mathbf{J} = 
    V^2 
    \begin{pmatrix}
        \frac{1.24}{\epsilon_-}+\frac{1.24}{\epsilon_+} & \frac{0.1}{\epsilon_-}+\frac{0.83}{\epsilon_+} \\
        \frac{0.1}{\epsilon_-}+\frac{0.83}{\epsilon_+} & \frac{0.57}{\epsilon_-}+\frac{0.55}{\epsilon_+}
    \end{pmatrix}.
\end{equation}
For values of $\epsilon_-$ and $\epsilon_+$ compatible with the formation of a local moment, both eigenvalues $j_a$ are positive and of the same order of magnitude (see SM \cite{bib:SM}). This places the Kondo signal primarily in the left wing of $\bf 2$, as seen in the experiments, see Fig.\ref{fig:exp}.


To corroborate the results obtained above, we compute the orbital-resolved spectral functions by means of NRG \cite{krishna-murthy1980,bulla2008} on a two-channel Anderson model.
For $\bf 1$, we remove the $\alpha=4$ channel, as it has negligible magnetic coupling to the molecule. The hybridization parameter $\Gamma=\pi \rho V^2$ for each system is chosen so that the full width at half peak is of the order of the Kondo temperature $k_B T_K \simeq 10$\,K found in experiment (see Fig. \ref{fig:exp}). Details of the procedure can be found in the SM section~\ref{SM-sec:NRG}\cite{bib:SM}.

\begin{figure}
\centering
    \includegraphics[width=0.9\linewidth,angle=0]{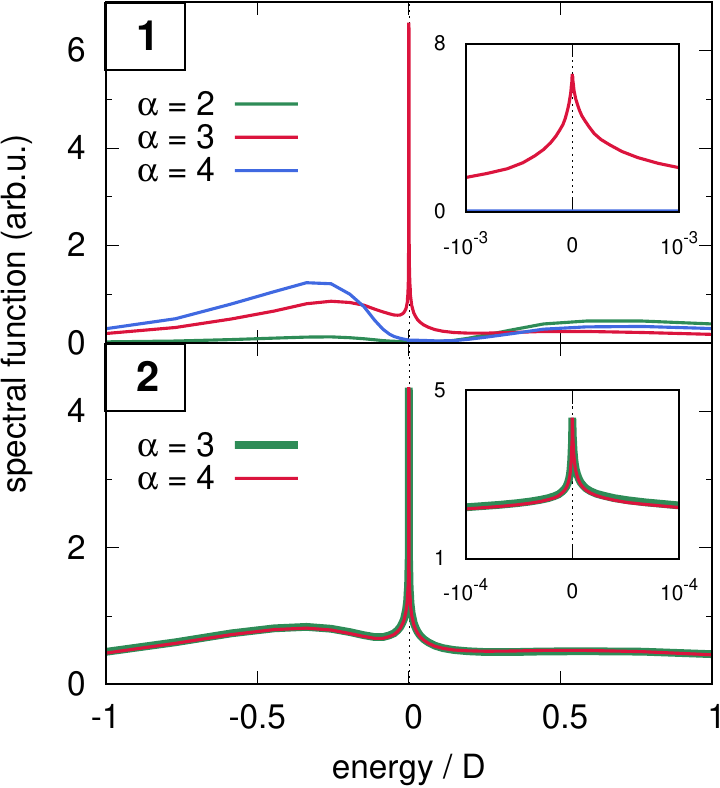}
\caption{Orbital-resolved spectral functions for $\bf 1$ (top panel) and $\bf 2$ (bottom panel). 
Insets: low-energy close-up.}
\label{fig:spectral}
\end{figure}

For $\bf 1$, the calculated spectral functions for $\epsilon_-=0.12$,
$\epsilon_+=0.29$, and a hybridization $\Gamma = 0.1$, all in units of 
the half-bandwidth $D=1\text{eV}$, are shown in Fig.
\ref{fig:spectral}. The appearance of a clear Kondo peak only in the channel
coupled to
the MO 3 supports our prediction that the effective AFM
coupling to the molecule happens primarily through channel 3. Thus, the conclusions from the previous section hold
even for the large hybridization values.
For $\bf 2$, the calculated spectral functions for $\epsilon_-=0.155$,
$\epsilon_+=0.3$, and $\Gamma=0.15$ are shown in Fig. \ref{fig:spectral}. Very
similar Kondo peaks  are found in the vicinity of both orbitals, as expected from
our previous discussion. 

In conclusion, we presented strongly correlated polyradical nanographenes showing coexistence of spatially delocalized Kondo and spin-flip excitation. We developed a perturbative approach of spin-flip excitation between various molecular multiplets, which enables us to identify the presence of antiferromagnetic screening channels. NRG calculations confirmed that AFM screening channels are responsible for multi-orbital Kondo screening in the polyradical nanographenes. 
We introduced the concept of the Kondo orbitals obtained from diagonalization of the magnetic exchange coupling matrix, which allow us to map a real-space localization of the Kondo resonance. We anticipate that the presented theoretical framework can be adopted for analysis of the emergence of the Kondo regime in other strongly correlated polyradical atomic and molecular systems.

\begin{acknowledgments}
A.C.-F., A.E. and M.B.-R. acknowledge grants No. IT-1527-22, funded by the Department of Education, Universities and Research of the Basque Government, and No. PID2019-103910GB-I00, funded by MCIN/AEI 10.13039/501100011033/. A.C.-F. acknowledges grant No. PRE2020-092046 funded by the Spanish MCIN.
D.S., A.P., O.S.,M.K. and P.J. acknowledge financial support from the CzechNanoLab Research Infrastructure supported by MEYS CR (LM2023051) and the GACR project no. 23-05486S. J. Lu acknowledges the support from MOE grants (MOE T2EP50121-0008 and MOE-T2EP10221-0005)  We acknowledge fruitful discussion with L. Veis and A. Matej. 
\end{acknowledgments}

\bibliography{bibliography.bib}

\makeatletter\@input{aux4_SM.tex}\makeatother

\end{document}


\title{Supplemental Material for ``Multi-orbital Kondo screening in strongly correlated polyradical nanographenes''}

\newcommand{\FISI}[0]{{Departamento de F\'{\i}sica, Universidad del Pa\'{\i}s Vasco UPV-EHU, 48080 Leioa, Spain}}
\newcommand{\QUIMI}[0]{{Departamento de Polímeros y Materiales Avanzados: F\'{\i}sica, Qu\'{\i}mica y Tecnolog\'{\i}a, Universidad del Pa\'{\i}s Vasco UPV-EHU, 20018 Donostia-San Sebasti\'an, Spain}}
\newcommand{\DIPC}[0]{{Donostia International Physics Center (DIPC),
20018 Donostia-San Sebasti\'an, Spain}}
\newcommand{\EHUQC}[0]{{EHU Quantum Center, Universidad del Pa\'{\i}s Vasco UPV-EHU, 48080 Leioa, Spain}}
\newcommand{\CFM}[0]{{Centro de F\'{\i}sica de Materiales CFM/MPC (CSIC-UPV/EHU), 20018 Donostia-San Sebasti\'an, Spain}}

\author{Aitor Calvo-Fern{\'a}ndez}
\affiliation{\FISI}
\affiliation{\DIPC} 

\author{Diego Soler-Polo}
\affiliation{Institute of Physics, Academy of Sciences of the Czech Republic, Cukrovarnicka 10, Prague 6, CZ 16200, Czech Republic}

\author{Andr{\'e}s Pinar Sol{\'e}}
\affiliation{Institute of Physics, Academy of Sciences of the Czech Republic, Cukrovarnicka 10, Prague 6, CZ 16200, Czech Republic}

\author{Shaotang Song}
\affiliation{Department of Chemistry, National University of Singapore, Singapore 117543, Singapore}

\author{Oleksander Stetsovych}
\affiliation{Institute of Physics, Academy of Sciences of the Czech Republic, Cukrovarnicka 10, Prague 6, CZ 16200, Czech Republic}

\author{Manish Kumar}

\affiliation{Institute of Physics, Academy of Sciences of the Czech Republic, Cukrovarnicka 10, Prague 6, CZ 16200, Czech Republic}

\author{Guangwu Li}
\affiliation{Department of Chemistry, National University of Singapore, Singapore 117543, Singapore}

\author{Jishan Wu}
\affiliation{Department of Chemistry, National University of Singapore, Singapore 117543, Singapore}

\author{Jiong Lu}
\affiliation{Department of Chemistry, National University of Singapore, Singapore 117543, Singapore}

\affiliation{Institute for Functional Intelligent Materials, National University of Singapore, 117544,
Singapore}

\author{Asier Eiguren}
\affiliation{\FISI}
\affiliation{\DIPC}
\affiliation{\EHUQC}

\author{Mar{\'i}a Blanco-Rey}
\affiliation{\QUIMI}
\affiliation{\DIPC}
\affiliation{\CFM}

\author{Pavel Jel\'{\i}nek}
\affiliation{Institute of Physics, Academy of Sciences of the Czech Republic, Prague, Czech Republic}

\affiliation{Regional Centre of Advanced Technologies and Materials, Czech Advanced Technology and Research Institute (CATRIN), Palacký University Olomouc,Olomouc 78371, Czech Republic}


\date{\today}

\maketitle

\section{Experiments}

The low-temperature SPM experiments were performed using a commercial low-temperature (1.2 K) STM/nc-AFM microscope Specs-JT with a Kolibri sensor (f $\approx$ 1 MHz) with a Standford Lock-in amplifier. High-resolution AFM and $\textit{{\it dI/dV}}$ maps were performed using a functionalized tip with a single carbon monooxide (CO) molecule. CO was dosed to the Au(111) at 4K. Typical CO coverages of two-three molecules per 20 $\times$ 20 nm$^2$ sample scan area were used. CO was picked up from the Au(111) by placing the tip on top of the molecule and increasing the sample bias in constant height STM mode until the CO functionalizes the tip ($\approx$ 200 pA at 130mV). The high-resolution AFM imaging was performed at constant height mode with bias voltage of 1 mV with oscillation amplitude of 50 pm. Lock-in parameters during $\textit{{\it dI/dV}}$ measurements: frequency: 723 Hz, modulation amplitude: 0.5 mV, time constant: 30 ms.


Nickelocene (NiCp$_2$) was dosed from a tantalum pocket kept at room temperature onto the prepared samples on Au(111) with submonolayer coverage at T $<$ 4K. To pick up the NiCp$_2$ on the STM tip, the sample was scanned at low current (10 pA STM setpoint and below 5 mV sample bias) until the nickelocene functionalizes the tip. Only stable NiCp$_2$ tips were used for IETS measurements. Additionally, the NiCp$_2$ tips were tested prior measurements by performing $\textit{d$^2$I/dV$^2$(z)}$ on Au(111) to observe the characteristic IETS peaks at $\approx$4 mV of the unperturbed NiCp$_2$ molecule. 

The fitting of the Kondo peaks was done with a Python script \cite{cahlik} that includes fitting routines with a Frota function \cite{frota1992shape} and a thermally broadened Fano function \cite{gruber2018kondo}, where an operator is defined for the convolution of the Fano line-shape with a derivative of Fermi-Dirac distribution. The fitting was performed with Model class in LMFIT, using Levenberg-Marquardt least-squares method.

\subsection{Magnetic origin of zero-bias peaks}
To demonstrate the magnetic origin of zero energy peaks observed in STS spectra, shown in Figure 1 in the main text, we carried out STS measurements with nickelocene-functionalized tip. Figure S1 displays the STS spectra measured along tip approach towards the sample, which clearly show a strong renormalization of the spin-excitation of the nickelocene tip at 5 meV along the tip approach due to exchange interaction between the spin of nickelocene tip and products {\bf 1} and {\bf 2}, respectively.

\begin{figure}[htp]

    \centering
    \includegraphics[width=17cm]{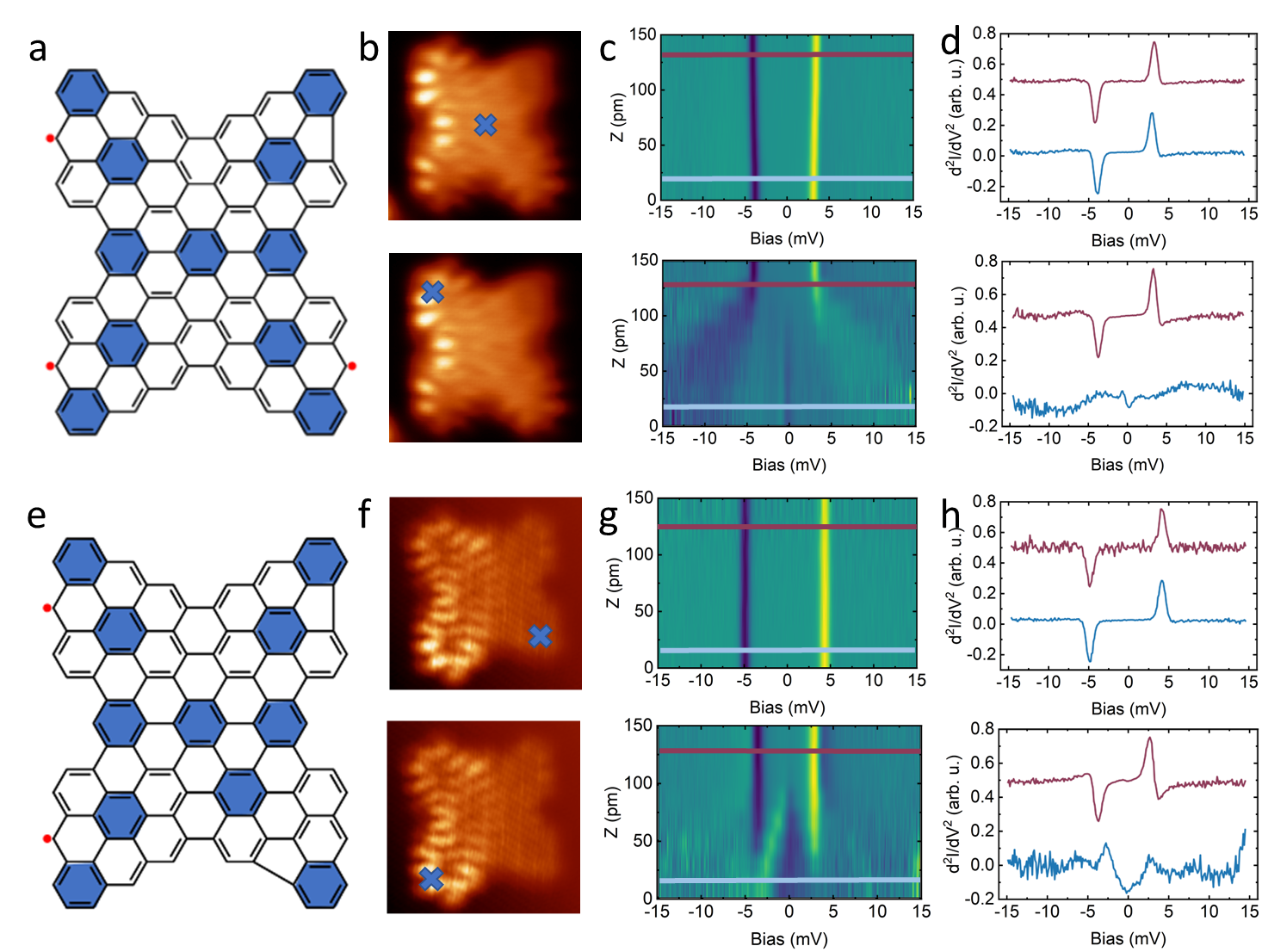}
     \caption{ \label{fig:IETS} {\bf $d^2I/dV^2$ IETS spectra of {\bf 1} and {\bf 2}  on Au(111) acquired with NiCp$_2$-functionalized tip.} (a, e) Chemical schemes of {\bf 1} and {\bf 2} molecules. Constant height STM images of {\bf 1} at 3.4 mV (b) and {\bf 2} at 1 mV (f), image size: 2.5 $\times$ 2.5 nm$^2$. (c, g) Experimental $d^2I/dV^2(z)$ IETS taken with NiCp$_2$-functionalized tip in the positions marked by blue crosses in top and bottom panels of (b, f) respectively. (d, h)  $d^2I/dV^2$ curves for selected heights in (c, g) marked with the horizontal lines. Red/blue colored lines in (c, g) correspond to red/blue spectra in (d, h). While the NiCp$_2$ $d^2I/dV^2(z)$ IETS remains unaltered when approaching to a non magnetic site of the molecules (b top, f top, it deforms on the egde providing a fingerprint of the local magnetic signal originated from the Kondo effect\cite{kondo1964resistance} (b bottom, f bottom).}
       \end{figure}
       

\begin{figure*}[!htbp]
\includegraphics[width=17cm]{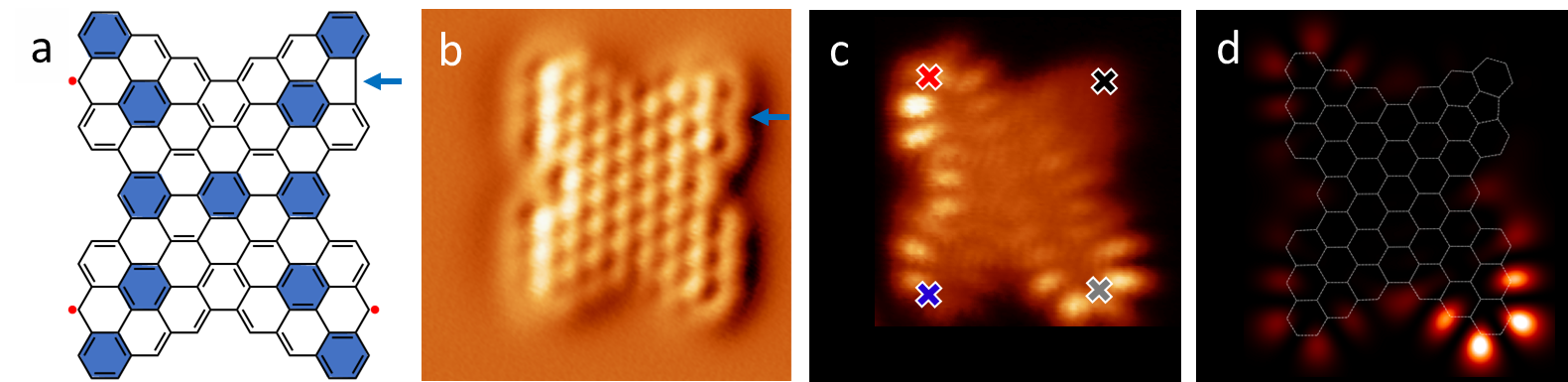}
\caption{{\bf Spin-excitation signal from the doublet ground state to the first excited quartet state of molecule 1}. a) scheme of the molecule, blue lines highlight the location of the defect. Simultaneously recorded high resolution CO tip AFM  (b) and ${\it dI/dV}$ map (c) images of the molecule {\bf 1} acquired at 24 meV, size: 3 $\times$ 3 nm$^2$. It should be noted that the signal on the left edge in the experimental dI/dV map is caused by the presence of the Kondo resonance located at the Fermi energy. Thus, the appearance of the signal on the left edge has nothing to do with spin excitation from the ground state to the first excited state. d) simulated spin excitation map obtained from the Natural Transition Orbitals for the spin flip operator between the ground state doublet and the first excited quartet state, see Figure S5.}
\label{fig:exp-SM-excDQ}
\end{figure*}

\section{Calculations}

\subsection{Natural transition orbitals}
To calculate Natural Transition Orbitals (NTO) orbitals between two states, $\vert \Psi_{-} \rangle $ and $\vert \Psi_{+} \rangle$  we construct for each $j$ and $k$ the matrix element:
  $$T_{j,k}  = \langle \Psi_-   \vert \hat{f}^\dagger_{j \downarrow} \hat{f}_{k \uparrow}     \vert \Psi_+  \rangle, $$
where, as above, $j$ is an index labeling molecular orbitals and $\hat{f}^\dagger_{j \sigma}$ are the creation operators for electrons in molecular spin-orbital $j \sigma.$ We then proceed, following Martin \cite{NTOs}, to diagonalize the matrix $TT^\dagger$ to obtain the NTOs. Finally, we use the PP-STM code \cite{Krej2017} to calculate corresponding {\it dI/dV} maps using s-like tip.

\begin{figure}[!htbp]
{\includegraphics[width=1.0\linewidth]{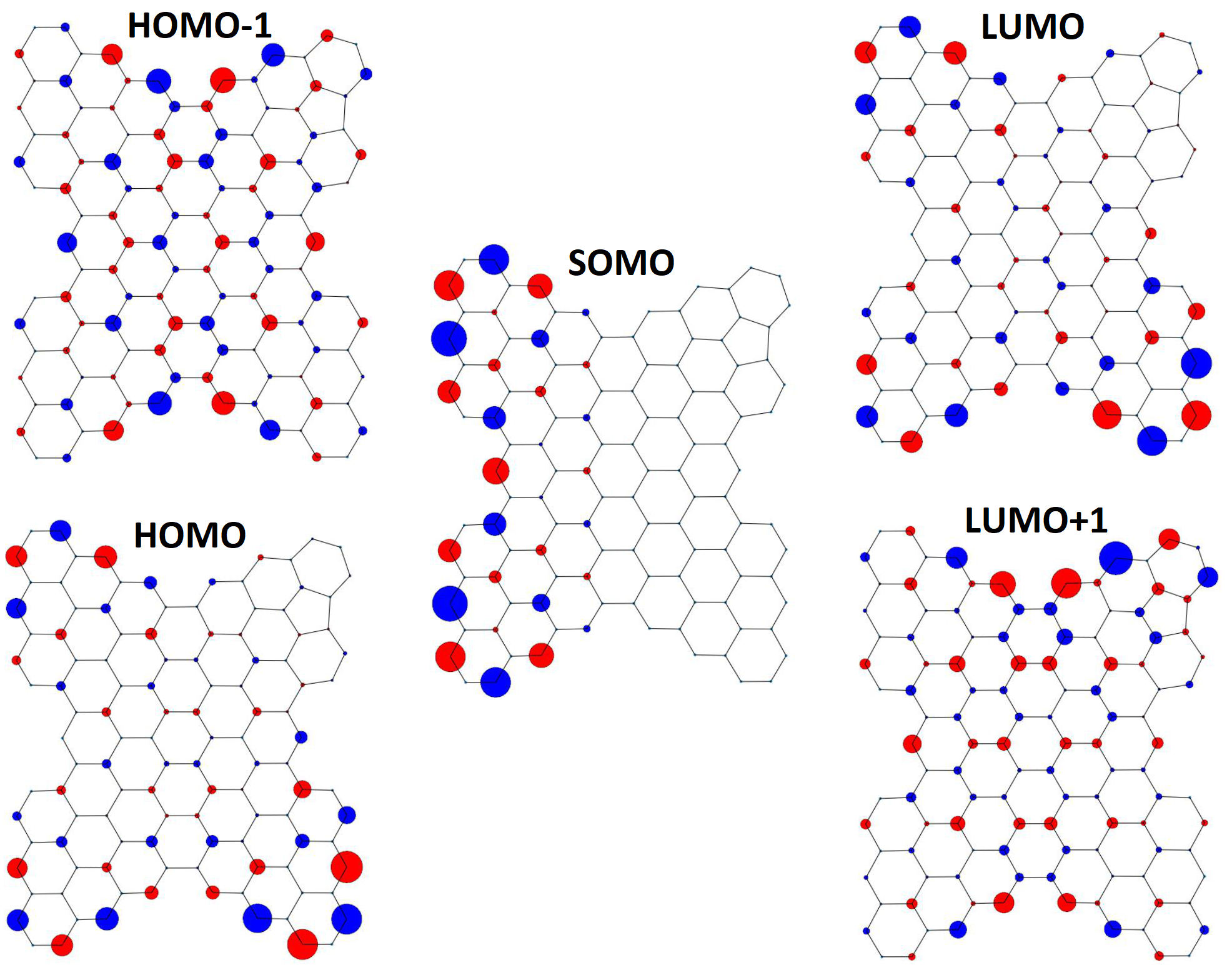}}
\caption{{\bf Single electron H{\"u}ckel orbitals used for the Hubbard-CAS calculations in product \textbf{1}}:  1 (HOMO-1), 2(HOMO), 3(SOMO), 4(LUMO), 5(LUMO+1). The numbering of orbitals is according to the Figure 2 in the main text.}
\label{fig: Molecular orbitals B3}
\end{figure}

\begin{figure}[!htbp]
{\includegraphics[width=1.0\linewidth]{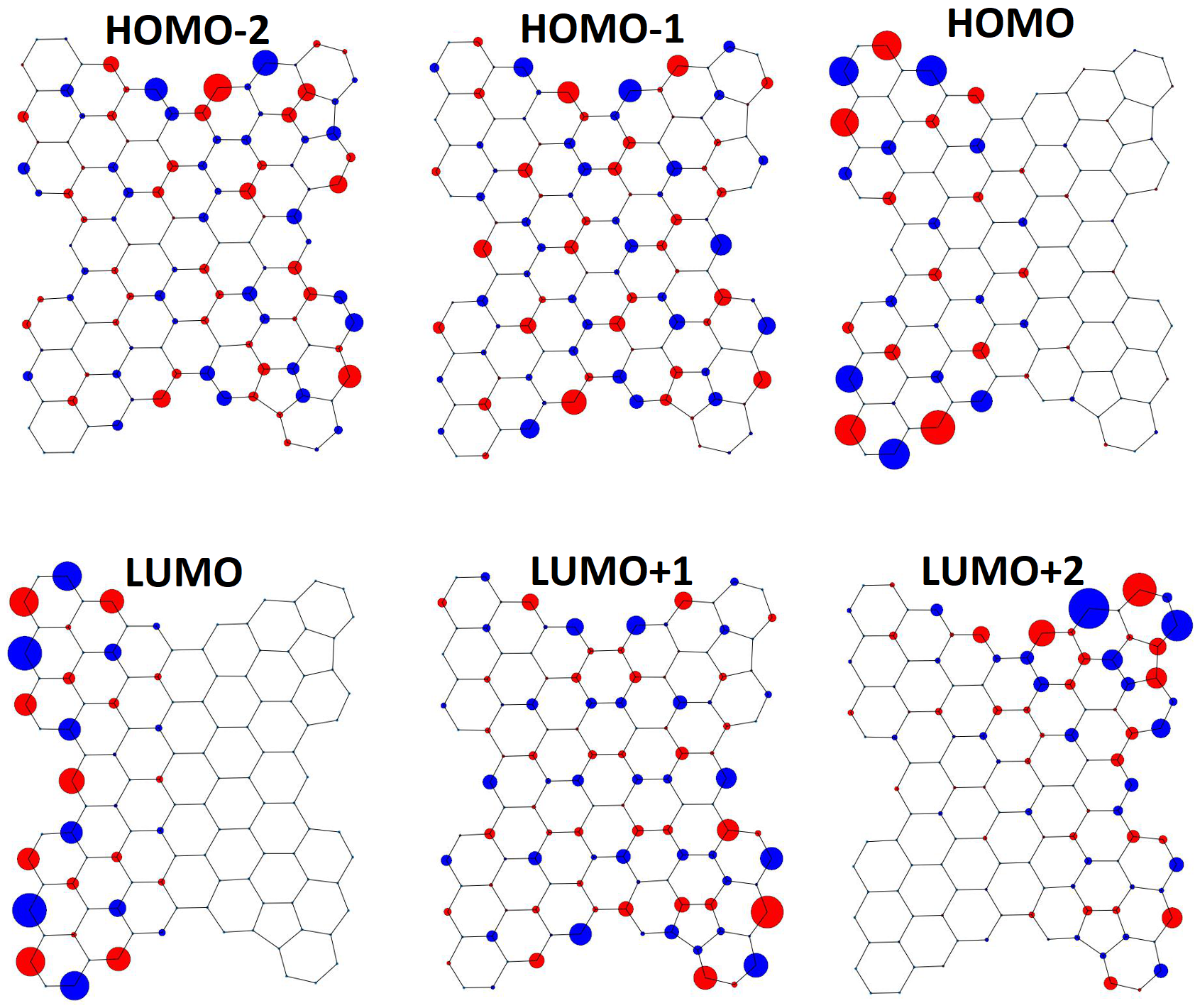}}
\caption{{\bf Single electron H{\"u}ckel orbitals used for the Hubbard-CAS calculations in product \textbf{2}}. First row, from left to right displays as follows : 1 (HOMO-2), 2(HOMO-1) and 3(HOMO). Second row, from left to right: 4 (LUMO), 5(LUMO+1), 6(LUMO+2). The numbering of orbitals is according to the Figure 2 in the main text.}
\label{fig: Molecular orbitals B2}
\end{figure}

\begin{figure}[!htbp]
{\includegraphics[width=1.0\linewidth]{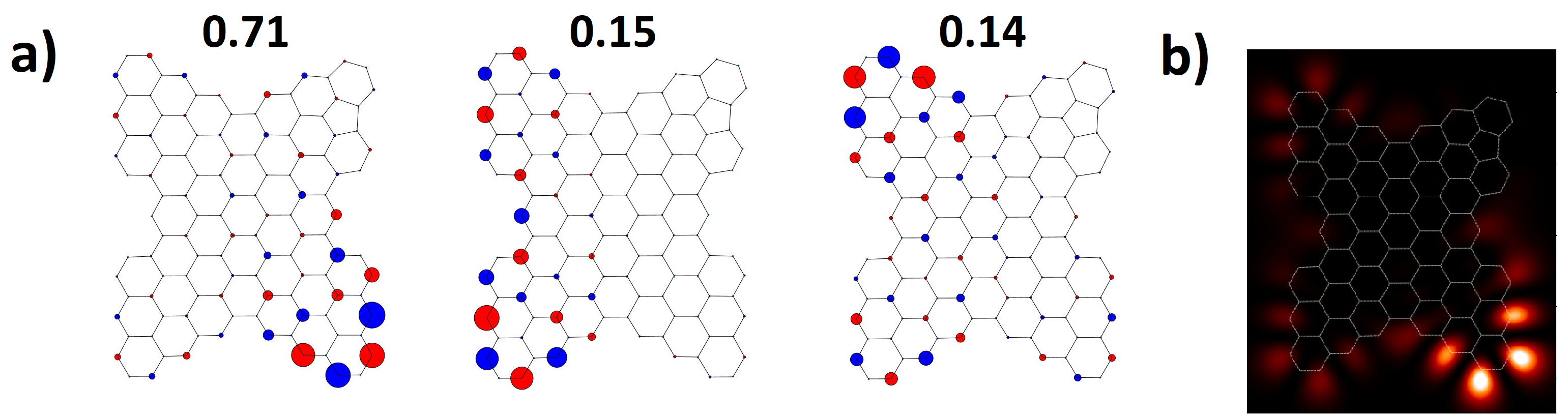}}
\caption{ {\bf Calculated NTO and {\it dI/dV} maps of the doublet to quartet spin excitation of molecule 1} a) Three dominant (from left to right) NTO orbitals for the doublet-quartet spin excitation of product \textbf{1}, each with their corresponding weights, b) resulting simulated {\it dI/dV} map obtained from these NTOs.}
\label{fig:NTO_B3_doublet2quartet}
\end{figure}

\begin{figure}[!htbp]
{\includegraphics[width=1.0\linewidth]{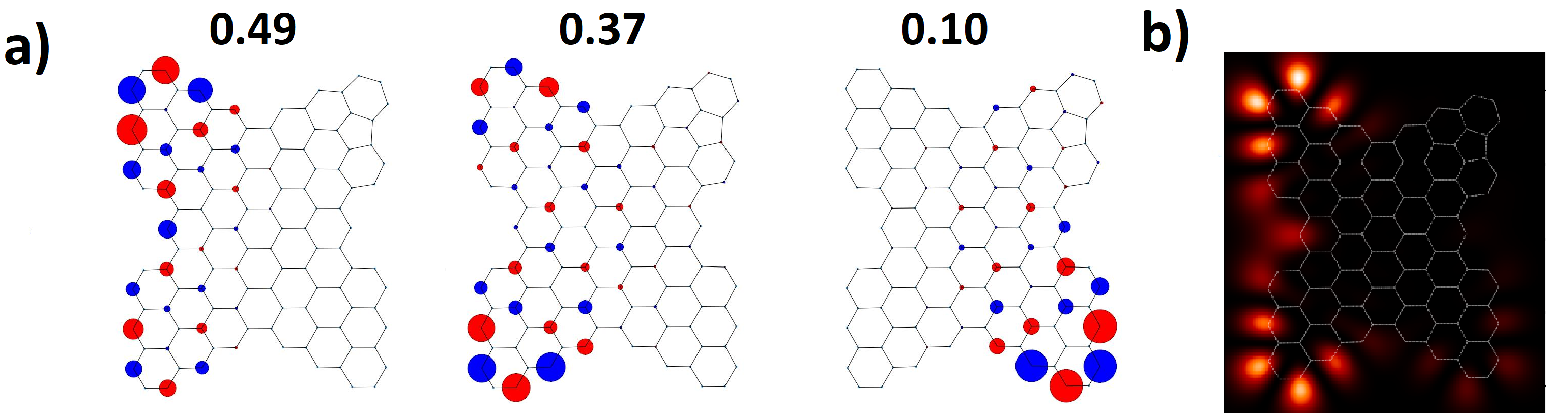}}
\caption{ {\bf Calculated NTO and {\it dI/dV} maps of the doublet to doublet spin flip of molecule 1.} a)Three dominant (from left to right) NTO orbitals for the spin flip between two degenerate ground state doublets that yields the Kondo effect in product \textbf{1} with their respective weights, b) simulated {\it dI/dV} map obtained from these orbitals }
\label{fig:NTO_B3_doublet2doublet}
\end{figure}

\begin{figure}[!htbp]
{\includegraphics[width=1.0\linewidth]{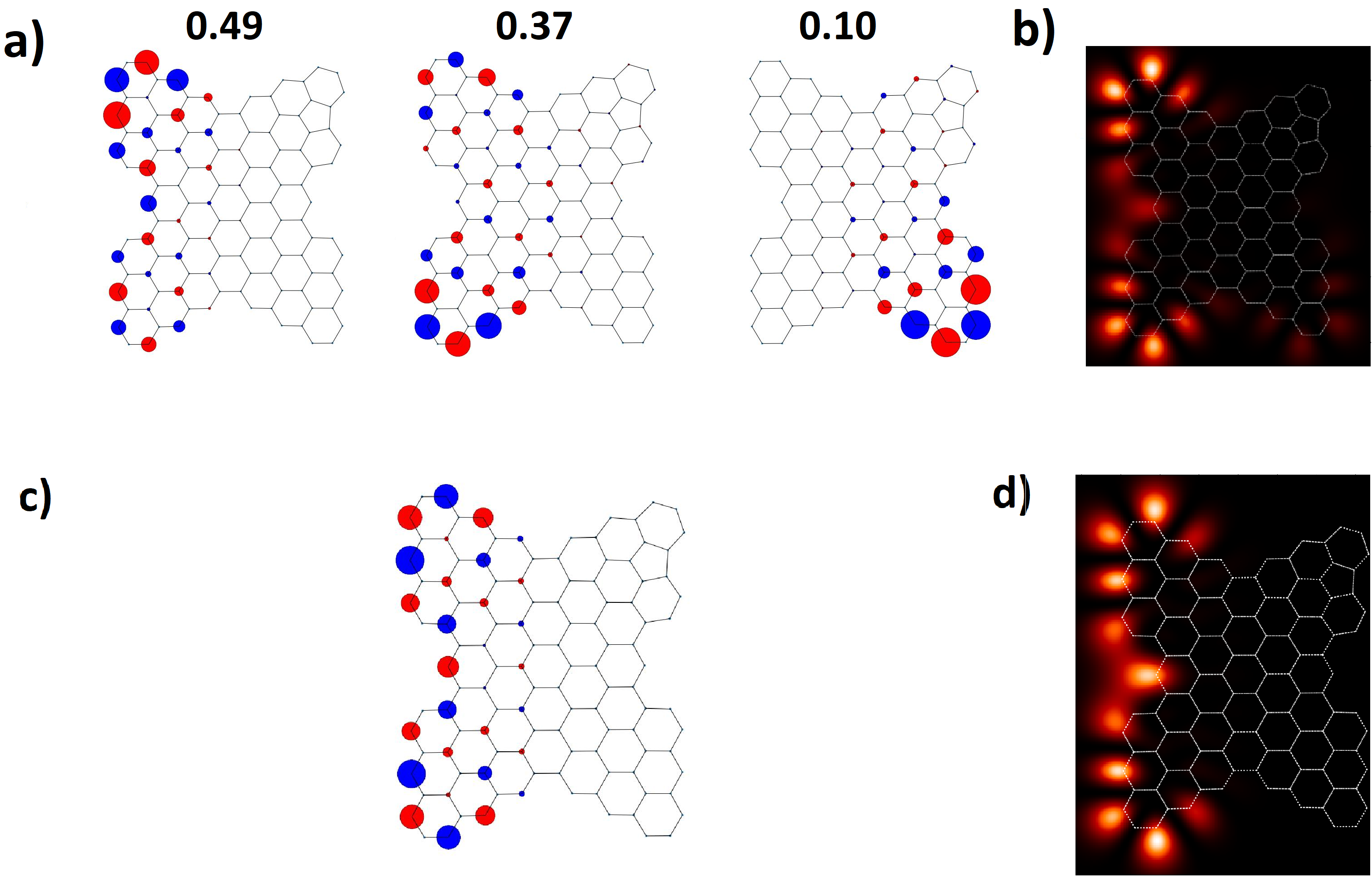}}
\caption{{\bf Comparison of {\it dI/dV} maps obtained from NTO and the Kondo channel for molecule 1.} a) NTO orbitals with their weigth and b) corresponding simulated {\it dI/dV} map. c) the Kondo channel obtained from the diagonalization of the $J$ matrix, and d) the corresponding simulated {\it dI/dV} map.}
\label{fig:compare_NTO_KO_B3}
\end{figure}

\begin{figure}[!htbp]
{\includegraphics[width=1.0\linewidth]{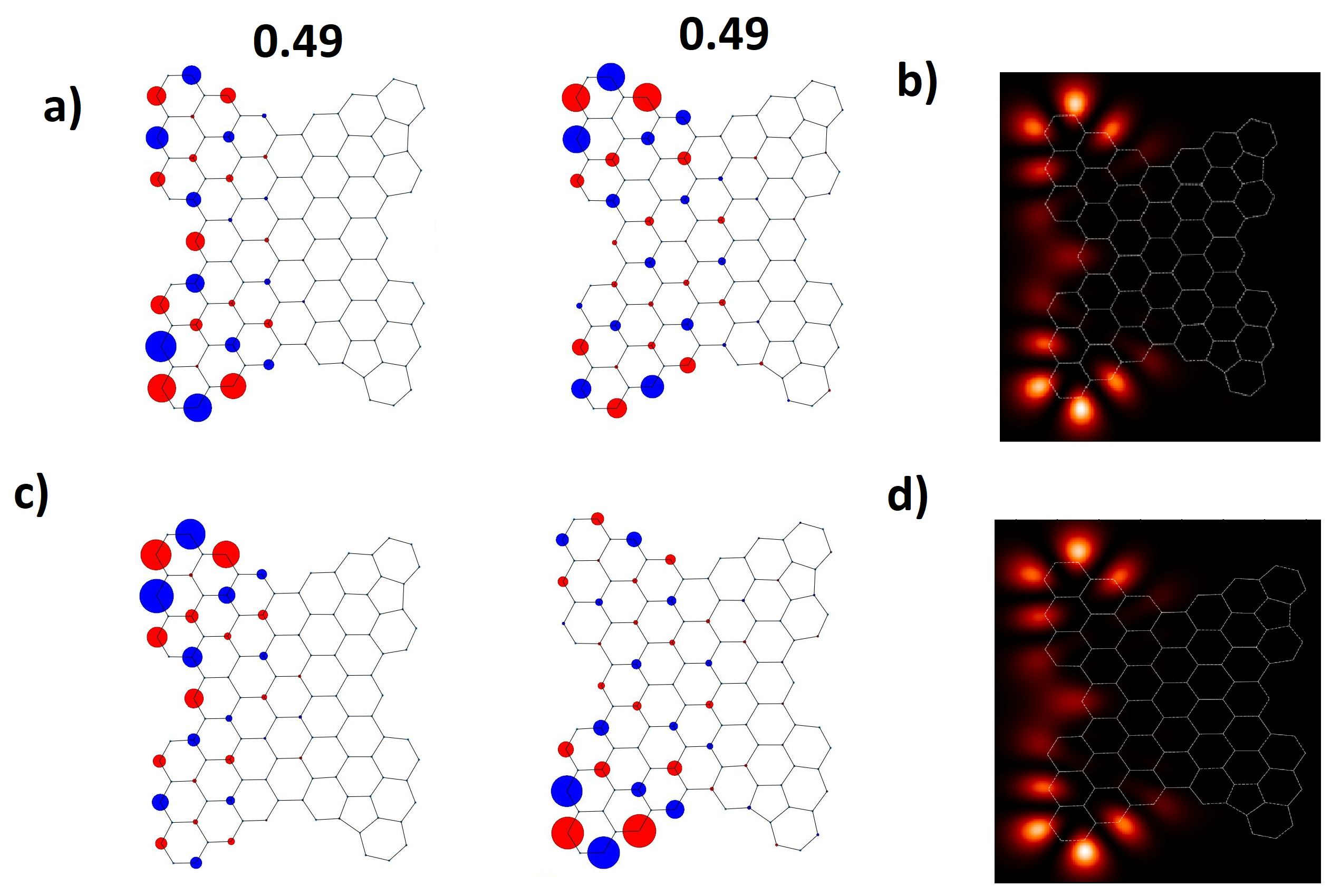}}
\caption{{\bf Comparison of {\it dI/dV} maps obtained from NTO and two Kondo channels for molecule 2.} a) NTO orbitals with their weight and b) corresponding simulated {\it dI/dV} map, c) two Kondo channels obtained from the diagonalization of the $J$ matrix and d) the corresponding simulated {\it dI/dV} map. }
\label{fig:compare_NTO_KO_B2}
\end{figure}

\begin{figure}[!htbp]
{\includegraphics[width=1.0\linewidth]{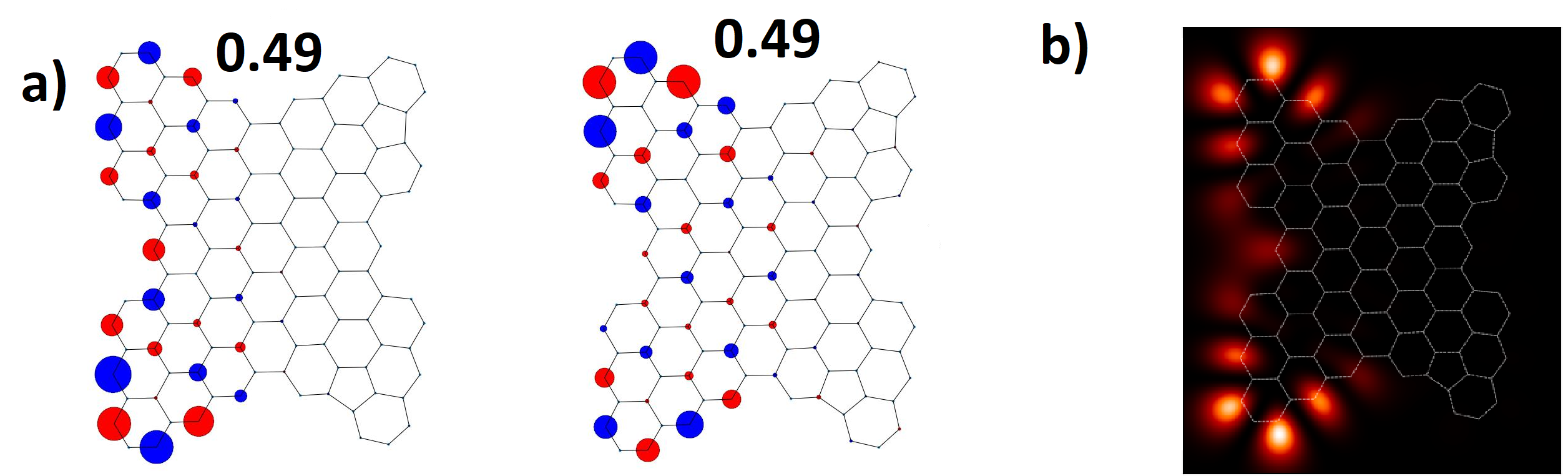}}
\caption{ {\bf Calculated NTO and {\it dI/dV} maps of the doublet to doublet spin flip of molecule 2.} a)Two dominant (from left to right) NTO orbitals for the spin flip between two degenerate ground state doublets that yields the Kondo effect in product \textbf{2} with their respective weights, b) simulated {\it dI/dV} map obtained from these two NTOs.}
\label{fig:NTO for the Kondo B2}
\end{figure}

\subsection{Spin-spin correlation of unpaired electrons}
We evaluated the spin correlations between the unpaired electrons in the degenerate ground states of products \textbf{1} and \textbf{2} as follows: 
$$\langle \Psi_0 \vert \Vec{\hat{S_j}} \cdot
\Vec{\hat{S_k}} \vert \Psi_0 \rangle - \langle \Psi_0 \vert
\Vec{\hat{S_j}}  \vert \Psi_0 \rangle \cdot \langle \Psi_0
\vert \Vec{\hat{S_k}}  \vert \Psi_0 \rangle,$$ 
where the indices $j$ and $k$ design the molecular orbitals. Here negative correlation represents antiferromagnetic coupling and positive ferromagnetic coupling between unpaired spins. We should note that the strong entanglement between three unpaired electrons is independent of the choice of the MO basis set.

The correlations between the unpaired spins are summarized in the Tables~\ref{spincorr1} and ~\ref{spincorr2}, where the molecular orbital "3" is the single-occupied zero-mode that survives on the butterfly after introducing the defect.

\begin{table}[htbp]
\centering
\begin{tabular}{|c|c|c|c|}
\hline
& 2 & 3 & 4    \\
\hline
2 &  0.050 &  -0.060 & 0.014   \ \\
3   &  -0.060 &  0.127 & -0.062    \ \\
4   &  0.014 &  -0.062 & 0.052  \ \\
\hline
\end{tabular}
\caption{Spin-spin correlations in \textbf{1} for the doublet ground state in basis of H{\"u}ckel molecular orbitals of the active space shown in Fig.\ref{fig: Molecular orbitals B3}. 
} 
\label{spincorr1}
\end{table}

\begin{table}[htbp]
\centering
\begin{tabular}{|c|c|c|c|}
\hline
& 3 & 4    \\
\hline
3 &  0.25 &  -0.24    \ \\
4   &  -0.24 &  0.25    \ \\
\hline
\end{tabular}
\caption{Spin-spin correlations in the $\bf 2$ for
    the doublet ground state in basis of H{\"u}ckel molecular
    orbitals of the active space shown in Fig.\ref{fig: Molecular orbitals B2}.
} 
\label{spincorr2}
\end{table}

\subsection{Electronic structure of product \textbf{1}: CASCI calculations}
Here we describe the wave-function  of the ground and excited states of {\bf 1} obtained from CASCI calculations for neutral ($N$) and charged ($N\pm1$) states.
\\ 

\textbf{Neutral molecule:}

The \textbf{1} wave-function in its doublet ground state in the $S_z = 1/2$ subspace is given by:

     \begin{align*}
        \vert M_0 \rangle \approx  0.70 \vert 2 \uparrow 0   \rangle + 0.47\vert 0 \uparrow 2 \rangle + 0.38\vert \uparrow \downarrow \uparrow \rangle +  0.21 \vert \uparrow \uparrow \downarrow   \rangle,
      \end{align*}
      whereas the $S_z=-1/2$ is obtained by applying the spin lowering operator.    

The quadruplet excited state \textbf{1}, located at 25 meV of the doublet ground state, has a somewhat simpler structure in the subspace $S_z = 1/2$:
             $$\vert M_1,S_z=1/2 \rangle \approx 0.57\vert \uparrow \uparrow \downarrow \rangle+0.57\vert \uparrow \downarrow \uparrow  \rangle+0.57\vert  \downarrow \uparrow \uparrow  \rangle. $$
             In the two equations above, orbitals 1 and 5 are dismissed, since their occupations are 2 and 0, and do not change.
             
             \textbf{Molecule with one extra electron:}
             The two main multiplets in the subspace $S_z=0$ are
                   $$\vert M^{+}_0,S_z=0 \rangle \approx 0.64\vert 2 2 \uparrow \downarrow 0 \rangle + 0.64\vert 2 2  \downarrow \uparrow 0 \rangle  + 0.13\vert 2 \uparrow 2 \downarrow 0 \rangle + 0.13\vert 2   \downarrow 2 \uparrow 0 \rangle  -0.14\vert 2  \uparrow \downarrow \uparrow \downarrow  \rangle  +0.14\vert 2   \downarrow \uparrow \downarrow \uparrow  \rangle,    $$
                    $$\vert M^{+}_1,S_z=0  \rangle \approx -0.72\vert 2 2 2 0 0 \rangle + 0.38 \vert 2 2 0 2 0 \rangle,$$
                    whereas in the $S_z=1$ subspace we have
                    $$\vert M^{+}_0, S_z = 1 \rangle \approx -0.91\vert 2 2 \uparrow \uparrow 0 \rangle  -0.18\vert 2 \uparrow 2 \uparrow 0 \rangle -0.19\vert 2 \uparrow \uparrow \uparrow \downarrow \rangle   -0.11\vert 2 2 \uparrow 0 \uparrow  \rangle - 0.11\vert 2 \uparrow \uparrow 2 0 \rangle.  $$

                    Their energies with respect to the ground state of the neutral molecule (CAS(5,5) for the Hubbard hamiltonian) are respectively: 487 meV and 491 meV. 
                    
             \textbf{Molecule with one less electron ($N-1$):}
             The two main multiplets are
             $$\vert M^{-}_0, S_z=0 \rangle \approx 0.70\vert 2  \uparrow \downarrow 0 0 \rangle + 0.70\vert 2   \downarrow \uparrow 0 0 \rangle,$$
              $$\vert M^{-}_0,S_z=0 \rangle \approx 0.87\vert 2  2 0 0 0 \rangle + 0.32\vert 2  0 2  0 0 \rangle + 0.30\vert 2  0 0  2 0 \rangle. $$ 
              Their energies with respect to the ground state of the neutral molecule (CAS(5,5) for the Hubbard hamiltonian) are respectively: -80 meV and -78 meV.  

\subsection{Electronic structure of product \textbf{2}: CASCI calculations}

The triplet and singlet many-body wavefunctions extracted from a CAS(6,6) are (here we only show occupancies of the orbitals 3 (HOMO) and 4(LUMO) because they are the most important ones within 96\% accuracy for the triplet ground state and the first excited singlet state):

\textbf{Neutral molecule:}

$$\vert M_{0},S_z=0 \rangle  \approx 0.7 \vert \uparrow \downarrow \rangle + 0.7 \vert \downarrow \uparrow \rangle. $$
In the subspace $S_z=1$ it is simply to 96\% of the Slater determinant:
$$\vert M_{0},S_z=1 \rangle  \approx \vert \uparrow \uparrow \rangle. $$

The excited singlet $\vert M_{1} \rangle$, on the other hand, is simply:
$$\vert M_{1} \rangle  \approx 0.7 \vert 2 0 \rangle + 0.7 \vert 0 2 \rangle $$
(here the occupation numbers always refer to the orbitals 3(HOMO) and 4(LUMO)).

 \textbf{Molecule with one extra electrons:}
             The two main multiplets in the subspace are
             $$\vert M_{0}^{+}, S_z = 1/2 \rangle  \approx 0.78 \vert 2 \uparrow \rangle +  0.52\vert \uparrow 2   \rangle,   $$
$$\vert M_{1}^{+},S_z = 1/2 \rangle  \approx -0.54 \vert 2 \uparrow \rangle +  0.81\vert \uparrow 2   \rangle.   $$
The energies of these multiplets with respect to the neutral Ground State are: 2.37eV and 2.41 eV

             \textbf{Molecule with one less electron:}
             The two main multiplets are
             $$\vert M_{0}^{-},S_z = 1/2 \rangle  \approx 0.78 \vert 0 \uparrow \rangle +  0.52\vert \uparrow 0   \rangle,   $$
$$\vert M_{1}^{-},S_z = 1/2 \rangle  \approx -0.54 \vert 0 \uparrow \rangle +  0.81\vert \uparrow 0   \rangle.   $$
The energies of these multiplets with respect to the neutral Ground State are: -1.91eV and -1.89eV. 


\section{Ionic Anderson Hamiltonian in the channel-energy representation} \label{sec:IAH}
In the discrete momentum space, the hybridization and conduction terms of the ionic Anderson Hamiltonian 
are
\begin{align}
    &\hat H_\text{con} 
    = 
    \sum_{\mathbf k, \sigma} 
    \epsilon_{\mathbf k} 
    \hat c ^\dagger _{\mathbf k \sigma}
    \hat c _{\mathbf k \sigma},
    \\
    &\hat H_\text{hyb}
    =
    \sum_{M_im_i}
    \sum_{M_fm_f}
    \sum_{\alpha,\mathbf k,\sigma}
    V_{\alpha \mathbf k} 
    \bra{M_fm_f} f^\dagger_{\alpha\sigma} \ket{M_im_i}
    \ket{M_fm_f}\bra{M_im_i}
    c_{\mathbf k\sigma}
    + \text{h.c.},
\end{align}
where $\epsilon_{\mathbf k}$ is the energy of the conduction electrons, $\hat c_{\mathbf k\sigma}$
($\hat c^\dagger_{\mathbf k\sigma}$) annihilates (creates) a conduction electron with momentum $\mathbf k$ 
and spin $\sigma$, and $V_{\alpha \mathbf k}$ is the amplitude of an electron tunneling from the MO $\alpha$ to 
a conduction state with momentum $\mathbf k$.
We begin by re-writing the sums in the continuous momentum space,
\begin{align}
    \sum_{\mathbf k} 
    \epsilon_{\mathbf k} 
    \hat c ^\dagger _{\mathbf k \sigma}
    \hat c _{\mathbf k \sigma}
    &\rightarrow
    \int d^3 k \epsilon_{\mathbf k} 
    \hat a^\dagger_{\mathbf k\sigma} 
    \hat a_{\mathbf k\sigma},
    \\
    \sum_{\mathbf k} V_{\alpha k} \hat a_{\mathbf k\sigma}
    &\rightarrow 
    \nu^\frac{1}{2}\int d^3 k 
    V_{\alpha\mathbf k}
    \hat a _{\mathbf k\sigma}.
\end{align}
where $a^{(\dagger)}_{\mathbf k\sigma}$ is the annihilation (creation) operator in the continuous $\mathbf k$-space, 
normalized as $\{a^\dagger_{\mathbf k\sigma},a_{\mathbf k'\sigma}\}=\delta(\mathbf k-\mathbf k')$, and $\nu$ is 
the density of $\mathbf k$ points. In order to change to the energy representation, we define the operators 
$\hat c^{(\dagger)}_{\epsilon \hat k \sigma} = 
|\nabla_{\mathbf k}\epsilon_{\mathbf k}|^{-\frac{1}{2}}a^{(\dagger)}_{\mathbf k\sigma}$, where $\hat k=\mathbf k/|\mathbf k|$
and the right hand side is evaluated at 
$\epsilon_{\mathbf k}=\epsilon$. This allows us to write the equations above as
\begin{align}
    &\int d^3 k \epsilon_{\mathbf k} 
    \hat a^\dagger_{\mathbf k\sigma} 
    \hat a_{\mathbf k\sigma}
    =
    \int d\epsilon 
    \int_{\epsilon}ds_{\mathbf k} 
    \epsilon
    \hat c^\dagger_{\epsilon \hat k \sigma}
    \hat c_{\epsilon \hat k \sigma}
    \\
    &\nu^\frac{1}{2}\int d^3 k 
    V_{\alpha\mathbf k}
    \hat a _{\mathbf k\sigma}
    =
    \int d\epsilon 
    \int_\epsilon \frac{ds_{\mathbf k}}{\sqrt{v(\mathbf k)}}
    V_{\alpha\mathbf k} 
    \hat c_{\epsilon \hat k \sigma},
\label{eq:Vc_kspace}
\end{align}
where $\int_\epsilon ds_{\mathbf k}$ represents an integral over the momentum space surface of
$\mathbf k$ points for which $\epsilon_{\mathbf k}=\epsilon$, and
$v^{-1}(\mathbf k):=v(\epsilon,\hat{\mathbf k})=\nu |\nabla_{\mathbf k}\epsilon|^{-1}_{\epsilon=\epsilon_{\mathbf k}}$.
Based on Eq. \ref{eq:Vc_kspace}, we define the channel-energy representation operators
as
\begin{equation}
    \hat c_{\epsilon\alpha\sigma} 
    = 
    \frac{1}{[\rho(\epsilon)]^\frac{1}{2} V_{\alpha}(\epsilon)} 
    \int_\epsilon \frac{ds_{\mathbf k}}{\sqrt{v(\mathbf k)}}
    V_{\alpha\mathbf k} 
    \hat c_{\epsilon \hat k\sigma},
\end{equation}
where 
\begin{equation}
    V_\alpha(\epsilon)
    =
    \left[ \frac{
        \int_\epsilon \frac{ds_{\mathbf{k}}}{v(\mathbf k)} |V_{\alpha\mathbf k}|^2
    }{
        \int ds_{\mathbf k}
    } \right]^\frac{1}{2}
\end{equation}
is the average of $|V_{\alpha\mathbf k}|^2$ over the $\epsilon_{\mathbf k}=\epsilon$ surface and 
\begin{equation}
    \rho(\epsilon) = \frac{
        \int_{\epsilon} \frac{ds_{\mathbf{k}}}{v(\mathbf k)} 
    }{
        \int_{\epsilon} ds_{\mathbf k}
    }
\end{equation}
is the density of states per unit of energy.
In general, these operators do not provide an orthonormal basis, since
\begin{equation}
    \{ \hat c^\dagger_{\epsilon\alpha\sigma} , \hat c_{\epsilon'\alpha'\sigma'} \}
    =
    \frac{\delta(\epsilon-\epsilon')\delta_{\sigma\sigma'}}{\rho(\epsilon)V_\alpha(\epsilon)V_{\alpha'}(\epsilon)}
    \int_{\epsilon_{\mathbf k}=\epsilon}\frac{ds_{\mathbf k}}{v(\mathbf k)}
    V^*_{\alpha \mathbf k} V_{\alpha'\mathbf k}.
\end{equation}
Nonetheless, we know from experimental measurements that that the hybridization between the MOs $\alpha$ 
is small, so we assume that 
\begin{equation}
    \int_\epsilon \frac{ds_{\mathbf k}}{v(\mathbf k)} 
    V_{\alpha\mathbf k}
    V_{\alpha'\mathbf k} 
    \approx 
    \rho(\epsilon)
    V_{\alpha}(\epsilon)^2 
    \delta_{\alpha\alpha'},
\end{equation}
which leaves us with an orthonormal basis that fulfils
\begin{equation}
    \{ c^\dagger_{\epsilon\alpha\sigma} , c_{\epsilon'\alpha'\sigma} \}
    =
    \delta(\epsilon-\epsilon')\delta_{\alpha\alpha'}\delta_{\sigma\sigma'}.
\end{equation}
Thus, in the channel-energy representation the full ionic Anderson Hamiltonian is
\begin{equation}
\begin{aligned}
    &\hat{\mathcal H} 
    = 
    \hat{H}_\text{mol} + \hat{H}_\text{con} + \hat{H}_\text{hyb},
    \\
    &\hat{H}_\text{mol} 
    = 
    \sum_M \sum_m \epsilon_M \ket{Mm}\bra{Mm},
    \\
    &H_{\text{con}} 
    =
    \sum_{\alpha\sigma}
    \int 
    \epsilon 
    \hat c^\dagger_{\epsilon\alpha\sigma}
    \hat c_{\epsilon\alpha\sigma}
    d\epsilon
    \\
    &H_{\text{hyb}}
    =
    \sum_{M_im_f}
    \sum_{M_fm_i}
    \sum_{\alpha\sigma}
    \int d\epsilon 
    [\rho(\epsilon)]^\frac{1}{2}
    V_{\alpha}(\epsilon) 
    \bra{M_fm_f} f^\dagger_{\alpha\sigma} \ket{M_im_i}
    \ket{M_fm_f}\bra{M_im_i}
    c_{\epsilon\alpha\sigma}
    +\text{h.c.},
\end{aligned}
\end{equation}
where we have introduced the $\hat H_\text{mol}$ containing the energies $\epsilon_M$ of each 
molecular multiplet $\ket{Mm}$
The Hamiltonian used in the main paper is obtained from considering the hybridization and the 
density of states as constants by performing the changes $V_\alpha(\epsilon)\rightarrow V$
and $\rho(\epsilon)\rightarrow\rho$.

\section{Effective Kondo coupling}\label{sec:Kondo}
    
\subsection{Schrieffer-Wolff transformation}
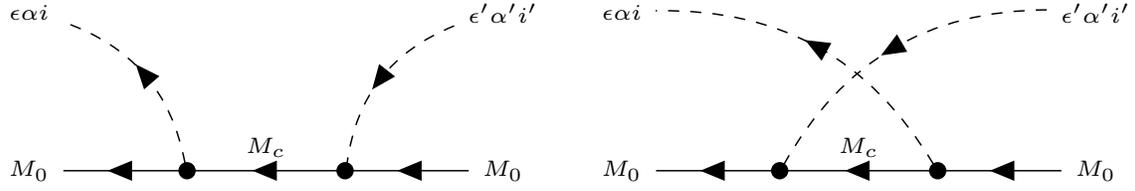
\begin{figure}
\scalebox{1.4}{
\begin{tikzpicture}
\begin{feynman}
\vertex (i1) {\tiny$M_0$};
\vertex [above=of i1] (e1) {\tiny$\epsilon\alpha i$};
\vertex [dot, right=of i1] (v1) {};
\vertex [dot, right=of v1] (v2) {};
\vertex [right=of v2] (i2) {\tiny$M_0$};
\vertex [above=of i2] (e2) {\tiny$\epsilon'\alpha' i'$};
\diagram* {
(e2) -- [charged scalar, bend right] (v2) ,
(v1) -- [charged scalar, bend right] (e1),
(i2) -- [fermion] (v2) -- [fermion, edge label'=\tiny$M_c$] (v1) -- [fermion] (i1),
};
\end{feynman}
\end{tikzpicture} }
\scalebox{1.4}{
\begin{tikzpicture}
\begin{feynman}
\vertex (i1) {\tiny$M_0$};
\vertex [above=of i1] (e1) {\tiny$\epsilon\alpha i$};
\vertex [dot, right=of i1] (v1) {};
\vertex [dot, right=of v1] (v2) {};
\vertex [right=of v2] (i2){\tiny$M_0$};
\vertex [above=of i2] (e2){\tiny$\epsilon'\alpha' i'$};
\diagram* {
(e2) -- [charged scalar, bend right] (v1),
(v2) -- [charged scalar, bend right] (e1),
(i2) -- [fermion] (v2) -- [fermion, edge label'=\tiny$M_c$] (v1) -- [fermion] (i1),
};
\end{feynman}
\end{tikzpicture}
}
\caption{Schematic Feynman diagrams for virtual scattering process through the intermediate
molecular multiplet $M_c$ with $N_c=N_0+1$ (left) and $N_c=N_0-1$ (right). }
\label{fig:feynman_schrieffer-wolff}
\end{figure}

The effective Kondo interaction for the ionic Anderson Hamiltonian  
to second order in perturbation theory takes the form
\begin{equation}
\label{eq:schriefferwolff}
\begin{aligned}
    \hat H_K 
    &= 
    \sum_{m_0,m_0'}
    \int_{-D}^D
    d\epsilon 
    \int_{-D}^D 
    d\epsilon'
    T(m_0,\epsilon\alpha i\leftarrow m_0',\epsilon'\alpha' i')
    \ket{M_0,m_0}\bra{M_0,m_0'}
    c^\dagger_{\epsilon\alpha i}
    c_{\epsilon'\alpha' i'}
\end{aligned}
\end{equation}
where $T(m_0,\alpha\epsilon i\leftarrow m_0',\alpha'\epsilon' i')$ is the second order scattering amplitude 
corresponding to virtual processes where an incoming electron with quantum numbers $\epsilon'$ (energy), 
$\alpha'$ (channel) and $i'$ (spin) is scattered off the molecule into a state with quantum numbers $\epsilon$,
$\alpha$ and $\sigma$, and the molecule spin changes its spin from $m_0'$ to $m_0$. This amplitude can be written 
as a sum of contributions from the intermediate excited multiplets,
\begin{equation}
\begin{aligned}
    T(m_0,\epsilon\alpha i\leftarrow m_0',\epsilon'\alpha' i')
    &=
    \sum_{M_c}
    T^{N_c}_{S_c}(m_0,\epsilon\alpha i\leftarrow m_0',\epsilon'\alpha' i')
\end{aligned}
\end{equation}
where $T^{N_c}_{S_c}(m_0,\epsilon\alpha i\leftarrow m_0',\epsilon'\alpha' i')$
is the amplitude for processes where the intermediate molecular state belongs to the $M_c$ multiplet,
which has occupation and spin quantum numbers $N_c,S_c$. The Feynman diagrams for such processes are given
schematically in Fig. \ref{fig:feynman_schrieffer-wolff}.
\par 
The general expressions for scattering through
an intermediate state with one more electron ($N_c=N_0+1$) and one less electron ($N_c=N_0-1$) are
\begin{equation}
\begin{aligned}
    T^{N_0+1}_{S_c}(m_0,\epsilon\alpha i\leftarrow m_0',\epsilon'\alpha' i')
    =&
    -\rho V^2
    \frac{
        \bra{M_0}|f_{\alpha}|\ket{M_c}
        \bra{M_c}|f^\dagger_{\alpha'}|\ket{M_0}
    }{
        \epsilon_{M_c}-\epsilon_{M_0}
    }
    \\
    &\times(\tfrac{1}{2},  i ;S_0, m_0+ i'- i|S_c,m_0+ i')
    (\tfrac{1}{2},  i';S_0, m_0|S_c, m_0+ i'),
\label{eq:Tplus}
\end{aligned}
\end{equation}
\begin{equation}
\begin{aligned}
    T^{N_0-1}_{S_c}(m_0,\epsilon\alpha i\leftarrow m_0',\epsilon'\alpha' i')
    =&
    \rho V^2
    \frac{
        \bra{M_0}|f^\dagger_{\alpha}|\ket{M_c}
        \bra{M_c}|f_{\alpha'}|\ket{M_0}
    }{
        \epsilon_{M_c}-\epsilon_{M_0}
    }
    \\
    &\times 
    (\tfrac{1}{2},  i ;S_c, m_0- i'|S_0,m_0- i'+ i)
    (\tfrac{1}{2},  i';S_c, m_0- i'|S_0, m_0),
\label{eq:Tminus}
\end{aligned}
\end{equation}
where we have used the  Wigner-Eckart theorem to decompose the excitation matrix elements as
\begin{equation}
    \bra{M_1,m_1}f^\dagger_{\alpha i}\ket{M_2,m_2}
    =
    \bra{M_1}|f^\dagger_\alpha|\ket{M_2}
    ( \frac{1}{2},  i; S_2, m_2 | S_1, m_1)^*,
\end{equation}
where $\bra{M_1}|f^\dagger_\alpha|\ket{M_2}$ is the reduced matrix element with respect to spin
symmetry. The Clebsch-Gordan coefficients needed to evaluate the terms in Eqs. \ref{eq:Tplus} and
\ref{eq:Tminus} are
\begin{align}
    (\tfrac{1}{2},\pm\tfrac{1}{2};S,M|S+\tfrac{1}{2},M\pm\tfrac{1}{2}) 
    & = 
    \sqrt{\frac{1}{2}\left(1\pm\frac{M\pm\frac{1}{2}}{S+\tfrac{1}{2}}\right)}
    \\
    (\tfrac{1}{2},\pm\tfrac{1}{2};S,M|S-\tfrac{1}{2},M\pm\tfrac{1}{2}) 
    & =
    \mp \sqrt{\frac{1}{2}\left(1\mp\frac{M\pm\tfrac{1}{2}}{S+\tfrac{1}{2}}\right)}.
\end{align}
\par
We begin by evaluating $T^{N_0+1}_{S_0-\frac{1}{2}}(m_0,\epsilon\alpha i\leftarrow m_0',\epsilon\alpha' i')$,
which yields
\begin{equation}
\label{eq:Tplusexpanded}
\begin{aligned}
    T^{N_0+1}_{S_0-\frac{1}{2}}(m_0,\epsilon\alpha i & \leftarrow m_0',\epsilon'\alpha' i')
    =
    -\rho V^2
    \frac{
        \bra{M_0}|f_{\alpha}|\ket{M_c}
        \bra{M_c}|f^\dagger_{\alpha'}\ket{M_0}
    }{
        \epsilon_{M_c}-\epsilon_{M_0}
    }
    \\
    &\times\begin{cases}
        &-\frac{1}{S_0+\frac{1}{2}}
        \frac{1}{2}
        m_0
        ,\text{ if } i= i'=\frac{1}{2},m_0=m_0',
        \\
        &+\frac{1}{S_0+\frac{1}{2}}
        \frac{1}{2}
        m_0
        ,\text{ if } i= i'=-\frac{1}{2},m_0=m_0',
        \\
        &-\frac{1}{S_0+\frac{1}{2}}
        \frac{1}{2}\sqrt{S_0(S_0+1)-m_0(m_0-1)}
        ,\text{ if } i=\frac{1}{2}, i'=-\frac{1}{2},m_0=m_0'-1,
        \\
        &-\frac{1}{S_0+\frac{1}{2}}
        \frac{1}{2}\sqrt{S_0(S_0+1)-m_0(m_0+1)}
        ,\text{ if } i=-\frac{1}{2}, i'=\frac{1}{2},m_0=m_0'+1,
    \end{cases}
    \\
    &= 
    \rho V^2
    \frac{
        \bra{M_0}|f_{\alpha}|\ket{M_c}
        \bra{M_c}|f^\dagger_{\alpha'}\ket{M_0}
    }{
        (\epsilon_{M_c}-\epsilon_{M_0})(S_0+\frac{1}{2})
    }
    \frac{1}{2}
    S^a_{m_0 m_0'}
    \sigma^a_{ii'}
\end{aligned}
\end{equation}
where we have ignored the contributions to the scattering potential. In order to obtain the final expression
we have made use of the relations
\begin{align}
    &\hat S ^a  \sigma ^a _{ii'}
    =
    \frac{1}{2}(\hat S^+  \delta_{i,i+1} + \hat S^-  \delta_{i,i-1} )
    + \hat S^z  \sigma^z_{ii'},
    \\
    &S^+_{m_0 m_0'} 
    = 
    \sqrt{S_0(S_0+1)-m_0(m_0+1)}
    \delta_{m_0,m_0'+1}
    \\
    &S^-_{m_0' m_0} 
    = 
    \sqrt{S_0(S_0+1)-m_0(m_0-1)}
    \delta_{m_0,m_0'-1}
    \\
    &S^z_{m_0 m_0'} = m_0 \delta_{m_0,m_0'},
\end{align}
where $\sigma$ are the Pauli matrices and the index $a$ labels the components of the vectors, which are to
be summed over when repeated. Following the same procedure for the remaining $N_c,S_c$ combinations, we obtain
\begin{align}
    T^{N_0-1}_{S_0-\frac{1}{2}}(m_0,\epsilon\alpha i \leftarrow m_0',\epsilon'\alpha' i)
    &=
    \rho V^2
    \frac{
        \bra{M_0}|f^\dagger_{\alpha}|\ket{M_c}^*
        \bra{M_c}|f_{\alpha'}|\ket{M_0}^*
    }{
        (\epsilon_{M_c}-\epsilon_{M_0})S_0
    }
    \frac{1}{2}
    S^a_{m_0 m_0'}
    \sigma^a_{ii'},
    \\
    T^{N_0+1}_{S_0+\frac{1}{2}}(m_0,\epsilon\alpha i \leftarrow m_0',\epsilon'\alpha' i)
    &=
    - \rho V^2
    \frac{
        \bra{M_0}|f_{\alpha}|\ket{M_c}
        \bra{M_c}|f^\dagger_{\alpha'}|\ket{M_0}
    }
    {
        (\epsilon_{M_c}-\epsilon_{M_0})(S_0+\frac{1}{2})
    }
    \frac{1}{2}
    S^a_{m_0 m_0'}
    \sigma^a_{ii'}
    \\
    T^{N_0-1}_{S_0+\frac{1}{2}}(m_0,\epsilon\alpha i \leftarrow m_0',\epsilon'\alpha' i)
    &=
    -
    \rho V^2
    \frac{  
        \bra{M_0}|f^\dagger_{\alpha}|\ket{M_c}^*
        \bra{M_c}|f_{\alpha'}|\ket{M_0}^*
    }{
        (\epsilon_{M_c}-\epsilon_{M_0})(S_0+1)
    }
    \frac{1}{2}
    S^a_{m_0 m_0'}
    \sigma^a_{ii'}
\end{align}
Substituting these scattering amplitudes into Eq. \ref{eq:schriefferwolff} and introducing the operator
\begin{equation}
    \hat \psi _{\alpha i}
    =
    \int d\epsilon \rho^{\frac{1}{2}} \hat c_{\epsilon\alpha i},
\end{equation}
we can write the Kondo interaction as
\begin{equation}
    \hat H _\K
    = 
    \frac{J_{\alpha\alpha'}}{2}
    \hat S^a
    \cdot 
    \hat \psi ^\dagger _{\alpha i}
    \sigma^a_{ii'}
    \hat \psi _{\alpha' i'},
\end{equation}
where
\begin{align}
    &\hat S^a = \sum_{m_0,m_0'} S^a_{m_0,m_0'}\ket{M_0m_0}\bra{M_0m_0'},
    \\
    &J_{\alpha\alpha'}
    =
    \sum_{M_c}
    \frac{V^2}{\epsilon_{M_c}-\epsilon_{M_0}}
    C_{\alpha\alpha'}^{M_c}
    \\
    &C_{\alpha\alpha'}^{M_c}=
    \begin{cases}
    \frac{
        \bra{M_0}|f_{\alpha}|\ket{M_c}
        \bra{M_c}|f^\dagger_{\alpha}|\ket{M_0}
    }{
        (S_0+\frac{1}{2})
    }, &\text{if } N_c=N_0+1,S_c=S_0-\frac{1}{2},
    \\
    \frac{
        \bra{M_0}|f^\dagger_{\alpha}|\ket{M_c}^*
        \bra{M_c}|f_{\alpha}|\ket{M_0}^*
    }{
        S_0
    }, &\text{if } N_c=N_0-1,S_c=S_0-\frac{1}{2},
    \\
    -\frac{
        \bra{M_0}|f_\alpha|\ket{M_c}
        \bra{M_c}|f^\dagger_\alpha|\ket{M_0}
    }{
        (S_0+\frac{1}{2})
    }, & \text{if } N_c=N_0+1,S_c=S_0+\frac{1}{2},
    \\
    -\frac{
        \bra{M_0}|f^\dagger_\alpha|\ket{M_c}^*
        \bra{M_c}|f_\alpha|\ket{M_0}^*
    }{
        (S_0+1)
    }, & \text{if } N_c=N_0-1,S_c=S_0+\frac{1}{2}.
    \end{cases}
\label{eq:Ccoefficients}
\end{align}

\subsection{Scaling equations for the coupling constants}
In order to obtain the scaling equations for the coupling constants $J_{\alpha\alpha'}$ and
$j_a$, we apply the poor man's scaling method \cite{anderson1970} to the Kondo interaction derived in the previous section.
By integrating over second-order scattering processes where the intermediate
electron lies in the energy range $|\epsilon|\in[D-|\delta D|]$, we obtain the differential scaling 
$\delta H_\K$ of the Kondo interaction,
\begin{equation}
    \delta \hat H _K
    =
    \sum_{m_0,m_0'}
    \int_{-D}^D d\epsilon 
    \int_{-D}^D d\epsilon'
    \delta T(m_0,\epsilon\alpha  i\leftarrow m_0',\epsilon'\alpha'i')
    \ket{M_0,m_0}
    \bra{M_0,m_0'}
    \hat c ^\dagger _{ \epsilon\alpha i}
    \hat c _{ \epsilon'\alpha' i'},
\label{eq:deltaHK}
\end{equation}
where $\delta T(m_0,\epsilon\alpha  i\leftarrow m_0',\epsilon'\alpha'i')$
is the sum over second order 
scattering amplitudes involving the creation of an intermediate particle ($p$) with or hole ($h$)
with energy $|\epsilon|\in[D-\delta D,D]$: 
\begin{equation}
    \delta T(m_0,\epsilon\alpha  i\leftarrow m_0',\epsilon'\alpha'i')
    =
    \delta T^{(p)}(m_0,\epsilon\alpha i\leftarrow m_0',\epsilon'\alpha'i')
    +
    \delta T^{(h)}(m_0,\epsilon\alpha  i\leftarrow m_0',\epsilon'\alpha' i').
\end{equation}
The Feynman diagrams associated to these scattering amplitudes are given in Fig. \ref{fig:feynman_spin-flip}.
Evaluating them, we obtain 
\begin{equation}
\begin{aligned}
    &m_0,\epsilon\alpha  i \leftarrow m_0'',\epsilon''\alpha'',i'' \leftarrow m_0',\epsilon'\alpha'i'
    \\
    \delta T^{(p)}(m_0,\epsilon\alpha i\leftarrow m_0',\epsilon'\alpha'i')
    &=
    \frac{1}{2}
    \rho
    J_{\alpha\alpha''}
    S^a_{m_0 m_0''}
    \sigma^a_{ii''}
    \frac{\dD }{-D}
    \frac{1}{2}
    \rho
    J_{\alpha''\alpha'}
    S^a_{m_0'' m_0'}
    \sigma^b_{i''i'}
    \\
    &=
    -\frac{1}{4}
    \rho^2
    \left[J^2\right]_{\alpha\alpha'}
    (S^a S^b)_{m_0 m_0'}
    (\sigma^a \sigma^b)_{ii'}
    \frac{\dD}{D}
\end{aligned}    
\end{equation}
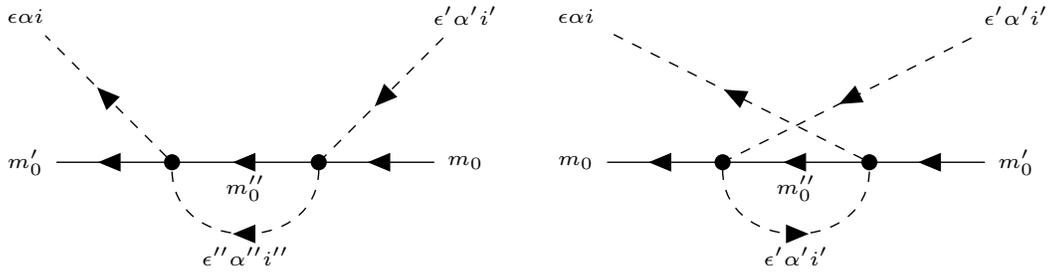
\begin{figure}
\scalebox{1.3}{
\begin{tikzpicture}
\begin{feynman}
\vertex (i1) {\tiny$m_0'$};
\vertex [above=of i1] (e1) {\tiny$\epsilon\alpha i$};
\vertex [dot, right=of i1] (v1) {};
\vertex [dot, right=of v1] (v2) {};
\vertex [right=of v2] (i2) {\tiny$m_0$};
\vertex [above=of i2] (e2) {\tiny$\epsilon'\alpha' i'$};
\diagram* {
(e2) -- [charged scalar] (v2) -- [charged scalar, half left, edge label=\tiny$\epsilon''\alpha''i''$] (v1) -- [charged scalar] (e1),
(i2) -- [fermion] (v2) -- [fermion, edge label=\tiny$m_0''$] (v1) -- [fermion] (i1),
};
\end{feynman}
\end{tikzpicture} }
\scalebox{1.3}{
\begin{tikzpicture}
\begin{feynman}
\vertex (i1) {\tiny$m_0$};
\vertex [above=of i1] (e1) {\tiny$\epsilon\alpha i$};
\vertex [dot, right=of i1] (v1) {};
\vertex [dot, right=of v1] (v2) {};
\vertex [right=of v2] (i2){\tiny$m_0'$};
\vertex [above=of i2] (e2){\tiny$\epsilon'\alpha' i'$};
\diagram* {
(e2) -- [charged scalar] (v1) -- [charged scalar, half right, edge label'=\tiny$\epsilon'\alpha' i'$] (v2) -- [charged scalar] (e1),
(i2) -- [fermion] (v2) -- [fermion,edge label=\tiny$m_0''$] (v1) -- [fermion] (i1),
};
\end{feynman}
\end{tikzpicture}
}
\caption{Feynman diagrams for virtual processes with an intermediate electron (left) and hole (left).}
\label{fig:feynman_spin-flip}
\end{figure}
for the scattering through intermediate states with an excited electron, 
where $\rho \dD$ is the number of particle states in the $[D-\dD,D]$ energy
range. For the scattering through an intermediate state with the incoming electron 
plus an electron-hole pair, we have
\begin{equation}
\begin{aligned}
    &m_0, \epsilon\alpha i 
    \leftarrow 
    m_0,\epsilon\alpha i, \epsilon''\alpha''i'', \epsilon'\alpha' i'
    \leftarrow 
    m_0',\epsilon'\alpha'i'
    \\
    \delta T^{(h)}(m_0,\epsilon\alpha i\leftarrow m_0',\epsilon'\alpha'i')
    &=
    -
    \frac{1}{2}
    \rho
    J_{\alpha''\alpha'}
    S^a_{m_0 m_0''}
    \sigma^a_{i''i'}
    \frac{\dD }{-D}
    \frac{1}{2}
    \rho
    J_{\alpha\alpha''}
    S^b_{m_0'' m_0'}
    \sigma^b_{ii''}
    \\
    &=
    \frac{1}{4}
    \rho^2
    \left[J^2\right]_{\alpha\alpha'}
    (S^a S^b)_{m_0 m_0'}
    (\sigma^b \sigma^a)_{ii'}
    \frac{\dD}{D}.
\end{aligned}    
\end{equation}
Thus, the total amplitude is
\begin{equation}
\begin{aligned}
    \delta T(m_0,\epsilon\alpha i\leftarrow m_0',\epsilon'\alpha'i')
    &=
    -\frac{1}{4}
    \rho^2
    \left[J^2\right]_{\alpha\alpha'}
    (S^a S^b)_{m_0 m_0'}
    [\sigma^a, \sigma^b]_{ii'}
    \frac{\dD}{D}
    \\
    &=
    -\frac{1}{4}
    \rho^2
    \left[J^2\right]_{\alpha\alpha'}
    (S^a S^b)_{m_0 m_0'}
    2i\epsilon_{abc}\sigma^c_{ii'}    
    \frac{\dD}{D}
    \\
    &=
    -\frac{1}{4}
    \rho^2
    \left[J^2\right]_{\alpha\alpha'}
    i\epsilon_{abc} 
    [S^a, S^b]_{m_0 m_0'}
    \sigma^c_{ii'}    
    \frac{\dD}{D}
    \\
    &=
    -\frac{1}{4}
    \rho^2
    \left[J^2\right]_{\alpha\alpha'}
    i\epsilon_{abc} 
    i\epsilon_{abd}
    S^d_{m_0 m_0'}
    \sigma^c_{ii'}    
    \frac{\dD}{D}
    \\
    &=
    \frac{1}{4}
    \rho^2
    \left[J^2\right]_{\alpha\alpha'}
    2\delta_{cd} 
    S^d_{m_0 m_0'}
    \sigma^c_{ii'}    
    \frac{\dD}{D}
    \\
    &=
    \frac{1}{2}
    \rho^2 
    \left[J^2\right]_{\alpha\alpha'}
    S^c_{m_0 m_0'}
    \sigma^c_{ii'}    
    \frac{\dD}{D}.
\end{aligned}
\end{equation}
Introducing this in Eq. \ref{eq:deltaHK}, we get
\begin{equation}
\begin{aligned}
    \delta \hat H_\K
    &=
    \sum_{m_0,m_0'}
    \int_{-D}^D
    d\epsilon
    \int_{-D}^D
    d\epsilon'
    \frac{1}{2}
    \rho^2 
    \left[J^2\right]_{\alpha\alpha'}
    \frac{\dD}{D}
    S^a_{m_0 m_0'}
    \hat c ^\dagger _{\epsilon\alpha  i}
    \sigma^a_{ii'}    
    \hat c _{\epsilon'\alpha'  i'}.
    \\ &=
    \frac{1}{2}
    \rho
    \left[J^2\right]_{\alpha\alpha'}
    \frac{\dD}{D}
    \hat S^c
    \hat \psi ^\dagger _{\alpha i}
    \sigma^c_{ii'}    
    \hat \psi _{\alpha' i'}.
\end{aligned}
\end{equation}
By comparison with Eq. \ref{eq:deltaHK}, we find
\begin{equation}
    \delta J_{\alpha\alpha'}
    =
    \rho [J^2]_{\alpha\alpha'}\frac{\dD}{D}.
\end{equation}
In the diagonal channel basis, this reduces to the simple expression
\begin{equation}
    \delta j_a
    =
    \rho j_a^2\frac{\dD}{D},
\end{equation}
Therefore, as the energy scale is lowered, the magnitude $|j_a|$ of the coupling will grow 
in antiferromagnetically coupled channels ($j_a>0$) until reaching the characteristic perturbative 
divergence and slowly
decrease in the ferromagnetically coupled channels ($j_a<0$). 

\subsection{Analysis of the coupling for $\bf 1$ and $\bf 2$}\label{Diagonalization}
In order to examine the effective coupling of $\bf 1$ and $\bf 2$ to the substrate in detail,
we diagonalize the Kondo coupling. This allows us 
to represent the coupling in an eigenstate basis where the molecule behaves as a local moment coupled
to several independent channels, meaning that conduction electrons in one such channel cannot be 
scattered to another channel. We call these new channels "eigenchannels" and we label them with the
letter $a$, keeping $\alpha$ for the channels that couple to each MO, which we call "MO channels". 
We shall determine (i) whether the coupling  to the eigenchannels is AFM or FM, 
(ii) the relative strength of these eigenchannels, and (iii) the projection of these eigenchannels 
onto the MO channels.
\par 
We choose the possible energy values $\epsilon_-$ and $\epsilon_+$ by approximating them as 
$\epsilon_-=\epsilon_-^0 -\mu$ and $\epsilon_+=\epsilon_+^0 +\mu$, where $\epsilon_-^0$ and 
$\epsilon_+^0$ are the energies of the excited multiplets when the molecule is free, and 
$\mu$ is the chemical potential introduced to model the coupling to the substrate. This
leaves us with $\mu$ as a parameter, which we allow to vary in the range where $\epsilon_-,\epsilon_+>0$,
ensuring that the ground state does not change.
\par
\begin{figure}[!htbp]
{\includegraphics[width=0.8\linewidth]{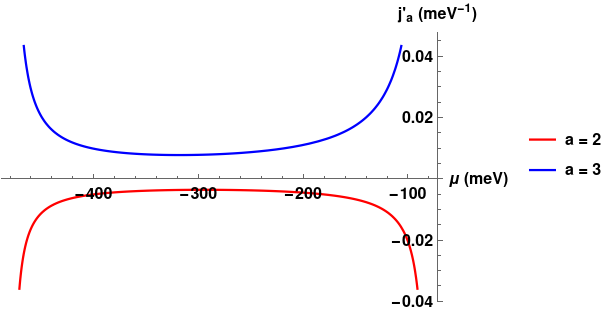}}
\caption{Eigenvalues $j'_2=V^{-2}j_2$ and $j'_3=V^{-2}j_3$ for $\bf 1$.}
\label{fig:B3_j2j3}
\end{figure}

\begin{figure}
    \begin{subfigure}{.49\textwidth}
        \centering
        \includegraphics[width=1.0\linewidth]{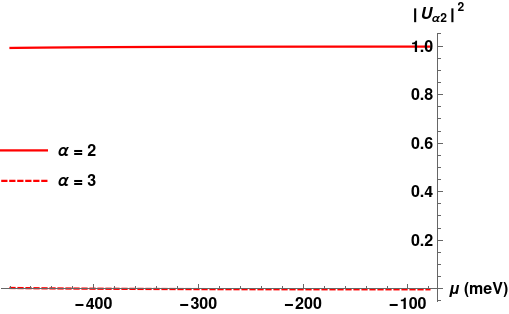}
        \caption{}
    \end{subfigure}
    \begin{subfigure}{.49\textwidth}
        \centering
        \includegraphics[width=1.0\linewidth]{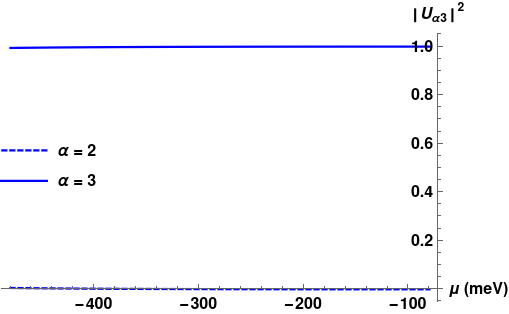}
        \caption{}
    \end{subfigure}
\caption{For $\bf 1$, MO channel mixings $|U_{\alpha a}|^2$ for the eigenchannels $a=2$ (a) and $a=3$ (b) and the primary (thick) and 
secondary (dashed) MO channels $\alpha=2,3$.}
\label{fig:B3_U}
\end{figure}

 
For $\bf 1$, using the data in Table I of the main paper, which corresponds to the channels $\alpha=2,3,4$ relevant for this molecule. we find that the coupling constant matrix is
\begin{equation}
    \mathbf{J}    
    =
    V^2
    \begin{pmatrix}
        -\frac{0.40}{\epsilon_-} - \frac{0.29}{\epsilon_+} & -\frac{0.06}{\epsilon_+} & 0 \\
        -\frac{0.06}{\epsilon_-} & \frac{1.13}{\epsilon_-}+\frac{0.51}{\epsilon_+} & 0 \\
        0 & 0 & -\frac{0.04}{\epsilon_{-}}-\frac{0.4}{\epsilon_+} 
    \end{pmatrix}
\end{equation}
We index the matrix elements $\alpha=2,3,4$ so that it matches the MO labeling. We immediately see
that the $\alpha=4$ MO channel is 
an eigenchannel with a FM coupling to the molecule. We discard it, as it is irrelevant at low temperatures.
In the upper block we find the eigenvalues $j_2$ and $j_3$,
which are shown Fig. \ref{fig:B3_j2j3}. Using the free molecule energies 
$\epsilon^0_-=-80$\,meV and $\epsilon^0_+=490$\,meV, we find that the eigenvalue $j_2$ 
is negative in the whole $\mu$ window, so the 
coupling to the $a=2$ eigenchannel is FM; $j_3$ is positive, so the coupling to $a=3$ is AFM. In Fig. \ref{fig:B3_U}
we show the mixing of the MO channels in the eigenchannels in terms of the quantities $|U_{\alpha a}|^2$, 
where $U$ are the diagonalizing matrices that fulfil $\mathbf{j}=U^\dagger \mathbf{J}U$. As can be seen in the graphs, the
eigenchannel 2 (3) couples almost exclusively to the MO channel 2 (3), hence the labeling of the eigenchannel.
The coupling to the secondary channel is of the order of $|U_{\alpha a}|\sim 10^{-3}$ for both eigenchannels.

\par
\begin{figure}[!htbp]
{\includegraphics[width=0.8\linewidth]{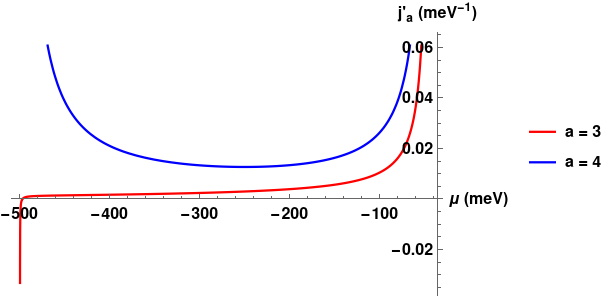}}
\caption{Eigenvalues $j'_3=V^{-2}j_3$ and $j_4'=V^{-2}j_4$ for $\bf 2$.}
\label{fig:B2_j3j4}
\end{figure}

\begin{figure}
    \begin{subfigure}{.49\textwidth}
        \centering
        \includegraphics[width=1.0\linewidth]{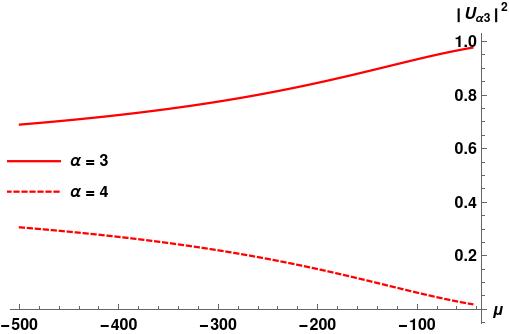}
        \caption{}
    \end{subfigure}
    \begin{subfigure}{.49\textwidth}
        \centering
        \includegraphics[width=1.0\linewidth]{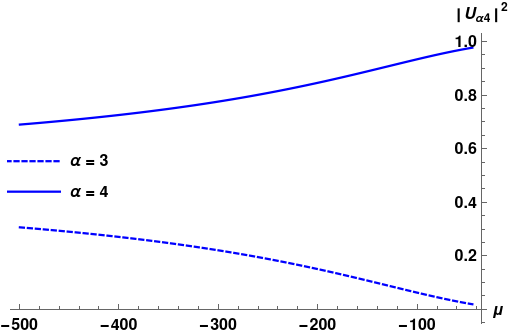}
        \caption{}
    \end{subfigure}
\caption{For $\bf 2$, MO channel mixings $|U_{\alpha a}|^2$ for the eigenchannels $a=3$ (a) and $a=4$ (b) and 
the primary (thick) and  secondary (dashed) MO channels $\alpha=3,4$.}
\label{fig:B2_U}
\end{figure}

 
For $\bf 2$, the coupling constant matrix calculated from the data on Table I of the main paper is given by
\begin{equation}
    \mathbf{J}
    = 
    V|^{2}
    \begin{pmatrix}
        \frac{1.24}{\epsilon_-}+\frac{1.24}{\epsilon_+} & \frac{0.1}{\epsilon_-}+\frac{0.83}{\epsilon_+} \\
        \frac{0.1}{\epsilon_-}+\frac{0.83}{\epsilon_+} & \frac{0.57}{\epsilon_-}+\frac{0.55}{\epsilon_+}
    \end{pmatrix}.
\end{equation}
Using the free molecule energies $\epsilon^0_-=-45$\,meV and $\epsilon_+^0=500$\,meV, we find the 
eigenvalues $j_3$ and $j_4$ shown in Fig. \ref{fig:B2_j3j4}. 
Except for the near vicinity of the lower limit $\mu=-500$\,meV, which is well into the mixed valence regime, 
both eigenvalues are positive, which means both eigenchannels
develop an AFM coupling with the molecular spin. The mixing of the MO channels in the eigenchannels is shown in 
Fig. \ref{fig:B2_U}. The mixing is larger for $\bf 2$ than for $\bf 1$ as a result of the relatively larger off-diagonal
entries of $\mathbf{J}$, but it is still possible to identify a primary MO channel for each eigenchannel in the whole $\mu$ region:
the eigenchannel 3 (4) has its largest component in the MO channel 3 (4).

\subsection{Relationship between the NTOs and the eigenchannels}\label{subsec:NTO-KO}
Using the orbital indices $\alpha$, we can write the transition matrix elements from which 
we compute the NTOs as
\begin{equation}
\begin{aligned}
    T^{m_0}_{\alpha\alpha'}
    &=
    \bra{M_0,m_0-1} 
    \hat f ^\dagger _{\alpha\downarrow} 
    \hat f _{\alpha'\uparrow} 
    \ket{M_0m_0}.
    \\
    &=
    \bra{M_0}| f^\dagger_\alpha f_{\alpha'} |\ket{M_0} 
    (1,-1;S_0,m_0|S_0,m_0-1),
\end{aligned}
\label{eq:Tdecomposition1}
\end{equation}
where we have applied the Wigner-Eckart theorem to decompose the transition matrix
element into a reduced part and a Clebsch-Gordan coefficient, taking into account 
that $f^\dagger_{\alpha\downarrow}f_{\alpha'\uparrow}$ is a tensor operator with 
total spin (irrep) $1$ and spin projection (partner number) $-1$. Since the transition
matrix elements in complete and reduced form for a given $m_0$ differ only in a
constant Clebsch-Gordan factor, they have the same eigenvectors and therefore they
give us the same NTOs.
\par 
We can also expand the transition matrix elements by introducing the identity operator 
$\mathbf I=\sum_{M,m}\ket{M,m}\bra{M,m}$, which results in 
\begin{equation}
\begin{aligned}
    T^{m_0}_{\alpha\alpha'}
    &=
    \sum_{M,m}
    \bra{M_0,m_0-1} 
    \hat f ^\dagger _{\alpha\downarrow} 
    \ket{Mm}
    \bra{Mm}
    \hat f _{\alpha'\uparrow} 
    \ket{M_0m_0}
    \\
    &=
    \sum_{M_c}
    \bra{M_0,m_0-1}
    \hat f ^\dagger _{\alpha\downarrow} 
    \ket{M_c,m_0-\tfrac{1}{2}}
    \bra{M_c,m_0-\tfrac{1}{2}}
    \hat f _{\alpha'\uparrow} 
    \ket{M_0m_0}
    \\
    &=
    \sum_{M_c}
    \bra{M_0}|
    \hat f ^\dagger _{\alpha} 
    |\ket{M_c}
    \bra{M_c}|
    \hat f _{\alpha'} 
    |\ket{M_0}
    (\tfrac{1}{2},+\tfrac{1}{2};S_c,m_0-\tfrac{1}{2}|S_0,m_0)
    (\tfrac{1}{2},-\tfrac{1}{2};S_c,m_0-\tfrac{1}{2}|S_0,m_0-1)
    \\
    &=
    S^-_{m_0-1,m_0}
    \sum_{M_c}
    \frac{
        \bra{M_0}|
        \hat f ^\dagger _{\alpha} 
        |\ket{M_c}
        \bra{M_c}|
        \hat f _{\alpha'} 
        |\ket{M_0}
    }{
        2 F(N_0-1,S_c,S_0) 
    }
    \\
    &=
    \bra{S_0}| S^- |\ket{S_0}
    (1,-1;S_0,m_0|S_0,m_0-1)
    \sum_{M_c}
    \frac{
        \bra{M_0}|
        \hat f ^\dagger _{\alpha} 
        |\ket{M_c}
        \bra{M_c}|
        \hat f _{\alpha'} 
        |\ket{M_0}
    }{
        2 F(N_0-1,S_c,S_0) 
    },
\end{aligned}
\label{eq:Tdecomposition2}
\end{equation}
where 
\begin{equation}
    F(N_c,S_c,S_0)=
    \begin{cases}
    (S_0+\frac{1}{2})
    , &\text{if } N_c=N_0+1,S_c=S_0-\frac{1}{2},
    \\
    S_0
    , &\text{if } N_c=N_0-1,S_c=S_0-\frac{1}{2},
    \\
    -(S_0+\frac{1}{2})
    , & \text{if } N_c=N_0+1,S_c=S_0+\frac{1}{2},
    \\
    -(S_0+1)
    , & \text{if } N_c=N_0-1,S_c=S_0+\frac{1}{2}
    \end{cases}
\label{eq:F}
\end{equation}
contains the spin factor and the sign appearing in Eq. \ref{eq:Ccoefficients}. The factors with 
$N_c=N_0+1$ appear when we compute the matrix elements with the creation and annihilation
operators commuted, 
\begin{equation}
    T^{m_0}_{\alpha\alpha'}
    =
    -\bra{M_0,m_0-1} f_{\alpha'\uparrow} f^\dagger_{\alpha\downarrow} \ket{M_0m_0}.
\end{equation}
\par 
Putting together Eqs. 
\ref{eq:Tdecomposition1} and \ref{eq:Tdecomposition2}, we arrive at 
the equivalent expressions
\begin{equation}
    \bra{M_0}| f^\dagger_\alpha f_{\alpha'} |\ket{M_0}
    =
    \bra{S_0}|S^-|\ket{S_0}
    \sum_{M_c}
    \frac{
        \bra{M_0}|
        \hat f ^\dagger _{\alpha} 
        |\ket{M_c}
        \bra{M_c}|
        \hat f _{\alpha'} 
        |\ket{M_0}
    }{
        2 F(N_0+1,S_c,S_0) 
    }
\end{equation}
and
\begin{equation}
    -\bra{M_0}| f_\alpha f ^\dagger _{\alpha'} |\ket{M_0}^*
    =
    \bra{S_0}|S^-|\ket{S_0}
    \sum_{M_c}
    \frac{
        \bra{M_0}|
        \hat f _{\alpha} 
        |\ket{M_c}^*
        \bra{M_c}|
        \hat f ^\dagger _{\alpha'} 
        |\ket{M_0}^*
    }{
        2 F(N_0-1,S_c,S_0) 
    }
\end{equation}
for the reduced matrix elements computed by introducing intermediate excited 
multiplets with positive and negative charge, respectively.
The reduced matrix element $\bra{S_0}|S^-|\ket{S_0}$ is constant, so we incorporate it
in a new transition matrix $\mathbf{\mathcal T}$ that contains only multiplet quantum numbers and gives us the
same NTOs as the original transition matrix $\mathbf T ^{m_0}$:
\begin{equation}
\begin{aligned}
    \mathcal T _{\alpha\alpha'}
    &:=
    \frac{ 
        2
        \bra{M_0}| f^\dagger_\alpha f_{\alpha'} |\ket{M_0}
    }{
        \bra{S_0}| S^- |\ket{S_0} 
    }
    =
    \sum_{M_c}^{N_c=N_0+1}
    \frac{
        \bra{M_0}|
        \hat f ^\dagger _{\alpha} 
        |\ket{M_c}
        \bra{M_c}|
        \hat f _{\alpha'} 
        |\ket{M_0}
    }{
        F(N_0+1,S_c,S_0) 
    }
    \\
    &=
    -\frac{ 
        2
        \bra{M_0}| f^\dagger_\alpha f_{\alpha'} |\ket{M_0}^*
    }{
        \bra{S_0}| S^- |\ket{S_0} 
    }
    =
    \sum_{M_c}^{N_c=N_0-1}
    \frac{
        \bra{M_0}|
        \hat f _{\alpha} 
        |\ket{M_c}^*
        \bra{M_c}|
        \hat f ^\dagger _{\alpha'} 
        |\ket{M_0}^*
    }{
        F(N_0+1,S_c,S_0) 
    }.
\end{aligned}
\end{equation}
\par 
To establish a connection between the NTOs and the eigenchannels, we write the coupling 
constant matrix as
\begin{equation}
    \mathcal J_{\alpha\alpha'}
    =
    V^{-2} J_{\alpha\alpha'}
    =
    \sum_{M_c}^{N_c=N_0-1}
    \frac{
        \bra{M_0}|
        \hat f ^\dagger _{\alpha} 
        |\ket{M_c}^*
        \bra{M_c}|
        \hat f _{\alpha'} 
        |\ket{M_0}^*
    }{
        (\epsilon_{M_c}-\epsilon_{M_0})
        F(N_0-1,S_c,S_0) 
    }
    +
    \sum_{M_c}^{N_c=N_0+1}
    \frac{
        \bra{M_0}|
        \hat f _{\alpha} 
        |\ket{M_c}
        \bra{M_c}|
        \hat f ^\dagger _{\alpha'} 
        |\ket{M_0}
    }{
        (\epsilon_{M_c}-\epsilon_{M_0})
        F(N_0+1,S_c,S_0) 
    }.
\end{equation}
We observe two main differences between $\mathbf{\mathcal T}$ and $\mathbf{\mathcal J}$:
(i) $\mathbf{\mathcal J}$ contains energy factors $(\epsilon_{M_c}-\epsilon_{M_0})^{-1}$ 
and (ii) $\mathbf{\mathcal T}$ can be equivalently obtained from projecting onto intermediate
states with either $N_c=N_0-1$ or $N_c=N_0+1$, while $\mathbf{\mathcal J}$ includes both types 
of terms. The similarity of the matrices is best observed in the case where all the multiplets
with $N_c=N_0 \pm 1$ particles are degenerate and have an energy $E=(\epsilon_{M_c}-\epsilon_{M_0})$
which yields
\begin{equation}
\begin{aligned}
    \mathcal J_{\alpha\alpha'}
    &=
    \frac{1}{E}
    \left[
    \sum_{M_c}^{N_c=N_0-1}
    \frac{
        \bra{M_0}|
        \hat f ^\dagger _{\alpha} 
        |\ket{M_c}^*
        \bra{M_c}|
        \hat f _{\alpha'} 
        |\ket{M_0}^*
    }{
        F(N_0-1,S_c,S_0) 
    }
    +
    \sum_{M_c}^{N_c=N_0+1}
    \frac{
        \bra{M_0}|
        \hat f _{\alpha} 
        |\ket{M_c}
        \bra{M_c}|
        \hat f ^\dagger _{\alpha'} 
        |\ket{M_0}
    }{
        F(N_0+1,S_c,S_0) 
    }
    \right]
    \\
    &=
    \frac{1}{E}\left[
    \mathcal T_{\alpha\alpha'}^* + \mathcal T_{\alpha\alpha'}^*
    \right]
    \\
    &=
    \frac{2}{E} \mathcal T _{aa}^*.
\end{aligned}
\end{equation}
This shows that in the limit where all the excited multiplets are degenerate, both matrices
contain equivalent information and therefore the same unitary transformation gives us the 
NTOs and the eigenchannels when applied to the orbital and conduction states, respectively. Furthermore, in Figs. \ref{fig:compare_NTO_KO_B3} and \ref{fig:compare_NTO_KO_B2} we show the numerical comparison between these orbitals and corresponding {\it dI/dV} maps. In the case of {\bf 1}, the agreement is good for the NTOs and the Kondo orbitals. The main difference in the {\it dI/dV} maps is the brighter contrast in the central part presented by the Kondo orbitals, which fits better to the experimental evidence (see Fig. 1 in the main text). In the case of {\bf 2}, we found no apparent difference between both {\it dI/dV} maps.

\section{NRG calculations}\label{sec:NRG}

\subsection{Method}
For the NRG calculations we use a discretized Anderson Hamiltonian with the conduction and hibridization terms
given by
\begin{equation}
\begin{aligned}
    & \hat H_\text{con} 
    = 
    \sum_{n=1}^\infty
    \tau_{n,\alpha} 
    \hat c^\dagger_{n,\alpha\sigma} 
    \hat c_{n+1,\alpha\sigma} 
    + 
    \text{h.c.},
    \\
    &\hat{H}_\text{hyb}
    = 
    \sum_{M_i,M_f}
    \sum_{m_i,m_f}
    \sqrt{2\Gamma/\pi}
    \bra{M_f,m_f}
    \hat f^\dagger_{\alpha \sigma}
    \ket{M_i,m_i}
    \ket{M_f,m_f}\bra{M_i,m_i}
    \hat{c}_{1,\alpha\sigma}
    + \text{h.c.},
\end{aligned}
\end{equation}
where $\Gamma=\pi\rho V^2$ is the constant hybridization; repeated indices are to be summed over. 
The term $\hat H_\text{con}$ represents the   
conduction channels as a semi-infinite tight-binding chains, which are obtained using the interleaved discretization
procedure \cite{yoshida1990}. Electrons in the $n$-th site of this chain
are annihilated (created) by the operators $\hat c^{(\dagger)}_{n,\alpha\sigma}$. Neighboring sites are coupled
by the hopping constants $\tau_{n,\alpha}$. The hybridization term $\hat H_\text{hyb}$
couples the molecule to the first site in the chain. For all the calculations we use a half-bandwidth of $D=1\text{eV}$.
\par 
We use the PointGroupNRG code \cite{calvofernandez2023} to compute the zero-temperature spectral functions and the thermodynamic curves (given below)
following the procedure presented in Ref. \cite{bulla2008}.  The spectral function, shown in Fig. 4 of the main paper, 
is defined as
\begin{equation}
    A_\alpha(\epsilon)
    =
    \sum_\sigma
    \sum_{E} 
    \left[
        |\bra{E} \hat f^\dagger_{\alpha\sigma} \ket{G}|^2 \delta(\epsilon-(\epsilon_E-\epsilon_G))
        +
        |\bra{E} \hat f_{\alpha\sigma} \ket{G}|^2 \delta(\omega+(\epsilon-\epsilon_G))
    \right],
\end{equation}
where $\ket{E}$ is an excited state, $\ket{G}$ is the ground state. The calculation is performed by choosing
for each NRG iteration $n$ a value $\epsilon_n=2\omega_n$ for which the spectral function is computed, 
where $\omega_n$ is the energy scale corresponding to the $n$-th step. The delta peaks are broadened using a 
Gaussian kernel with 
\begin{equation}
    \delta(\epsilon) \rightarrow \frac{1}{\eta_n\sqrt \pi} \exp\left[-\left(\epsilon/\eta_n\right)^2\right],
\end{equation}
where $\eta_n=\omega_n$ is chosen as the broadening factor. The results are averaged over 4 values of the twisting
parameter $z$ in order to improve the smoothness of the curves. For the thermodynamic functions, smooth curves are 
obtained by averaging over even and odd NRG steps. For both spectral and thermodynamic calculations, we use a 
discretization parameter $\Lambda=3$ and  a cutoff of 1000 multiplets.
\par

\subsection{Thermodynamic curves and Kondo temperatures}\label{thermoKondo}
To provide further insight into the Kondo effect, we show the thermodynamic curves for $\bf 1$ and
$\bf 2$. These have been computed using the same parameters $\pi\rho|V|^2$ and $\mu$ used in the 
calculation of the spectral functions.
\begin{figure}[!htbp]
\centering
\begin{subfigure}[b]{0.7\textwidth}
    \centering
    \includegraphics[width=\textwidth]{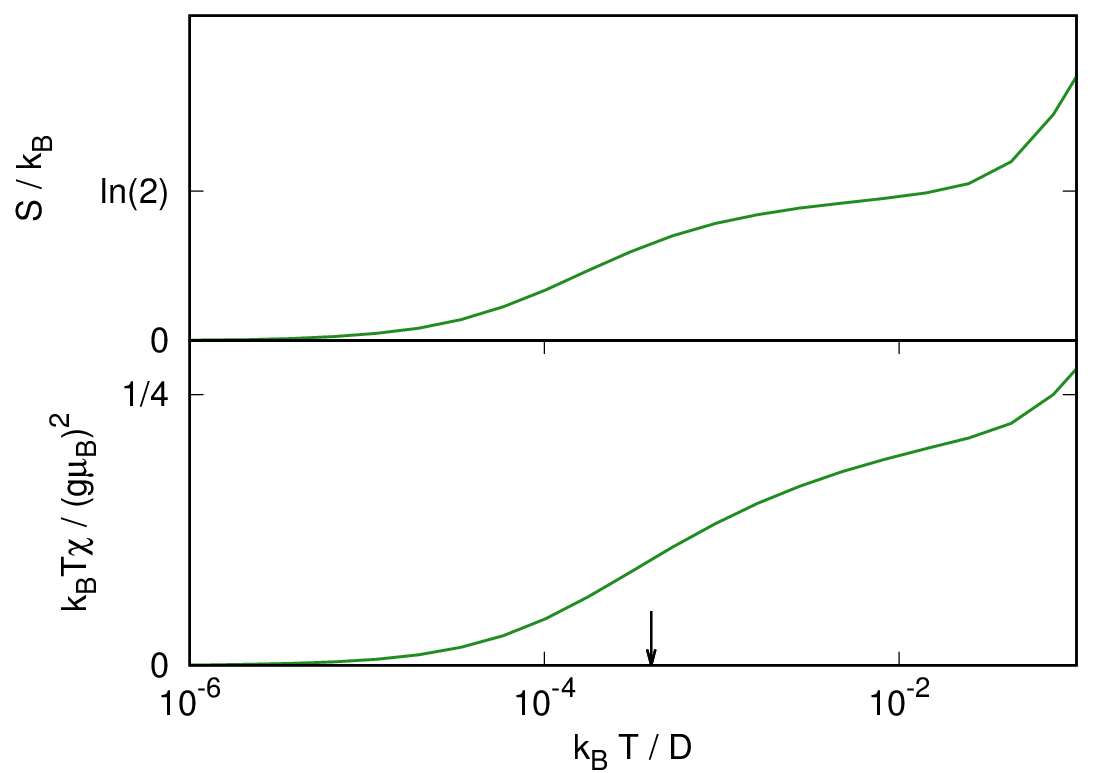}
    \caption{}
    \label{fig:B3_magentr}
\end{subfigure}
\hfill
\begin{subfigure}[b]{0.7\textwidth}
    \centering
    \includegraphics[width=\textwidth]{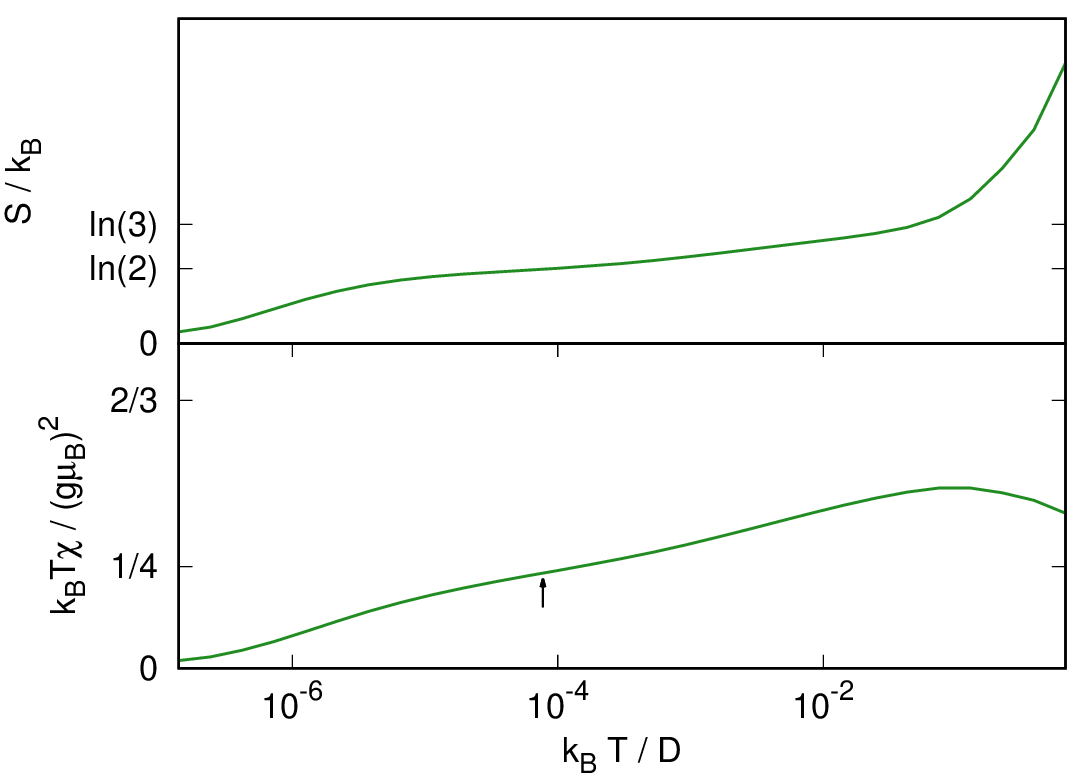}
    \caption{}
    \label{fig:B2_magentr}
\end{subfigure}
\caption{Calculated thermodynamic functions for molecules $\bf 1$ (a) and $\bf 2$ (b): 
entropy (top panel) and magnetic susceptibility (bottom panel). The parameters are
$\Gamma=0.1$\,eV for $\bf 1$, $\Gamma=0.15$\,eV for $\bf 2$, and $\mu=-0.2$\,eV for both. 
The arrow  in (a) indicates the Kondo temperature $k_B T_K = 4\times 10^{-4}$\,eV. The arrow 
in (b) indicates the separation between the two screening stages.}
\end{figure}
\par 
The calculated entropy and magnetic susceptibility curves for $\bf 1$ are shown in Fig. \ref{fig:B3_magentr}. These curves 
show that the system develops a local moment around $k_B T / D \sim 10^{-2}$, in the region
where the entropy and magnetic susceptibility are close to the local moment fixed point values 
$S/k_B=\ln(2)$ and $k_B T \chi / (g\mu_B)^2=1/4$. At lower temperatures this local moment is screened, 
with the full screening corresponding to the values $S=\chi=0$. The Kondo temperature $T_K$ obtained from the 
magnetic susceptibility, calculated as the point where $d\chi / dT |_{T=T_K}=0$, is $T_K\approx 5K$. This is close
to the value $T_K\approx 8K$ calculated from the full width at half-maximum $\Delta$ of the peak 
in the MO 3 spectral function, $k_B T_K = \Delta$.
Both values agree with the Kondo temperature obtained by fitting to the experimental results, which lies in the $T_K\in[1,10]K$
range.
\par 
Thermodynamic curves for $\bf 2$ are shown in Fig. \ref{fig:B2_magentr}. As the temperature is decreased,
the magnetic susceptibility curve starts approaching the $3/4$ value of the spin-$1$ local moment fixed point, 
which has an associated entropy $S/k_B=\ln(3)$. The hybridization rapidly causes this tendency to shift, screening the 
partially formed local moment. Two stages can be observed in the screening, separated by the middle inflexion point in 
the magnetic susceptibility curve, in agreement with the anistropy in the $j_a$ eigenvalues found in section \ref{Diagonalization}.
This is a characteristic feature of the spin-$1$ local moment asisotropic Kondo effect\cite{jayaprakash1981}.
\par 
It must be noted that, for both $\bf 1$ and $\bf 2$, the large
hybridization $\Gamma\sim\epsilon_\pm$ causes the system to evolve to the screened regime without fully developing 
the local moment. If the hybridization were smaller, we would observe a plateau close to the local moment
fixed point values for a large portion of  the temperature range. In principle, this would mean that the effective 
model is not a good approximation. Nevertheless, the results for the spectral functions agree with the analysis based on the 
effective Kondo model with respect to the nature of the magnetic coupling to the channels 2 and 3 (FM or AFM). Thus, 
the rationalization of the Kondo peaks based on the Kondo model can be considered as qualitatively correct.
\par 
\bibliography{bibliography.bib}